\documentclass[amsmath,twocolumn,superscriptaddress,prb,aps]{revtex4-1} 
\usepackage{amsthm,amsfonts,graphicx, color}
\usepackage{bm}
\usepackage{graphicx}
\usepackage{braket}
\usepackage{dcolumn}
\usepackage{bbm}
\usepackage{subfigure}
\usepackage{amssymb}
\usepackage[colorlinks,urlcolor=blue,linkcolor=blue,anchorcolor=blue,citecolor=blue]{hyperref}
\usepackage{wasysym}
\usepackage{color}
\stepcounter{secnumdepth}
\stepcounter{tocdepth}
\usepackage[normalem]{ulem}

\def\vr{\vec{r}}
\def\hx{\hat{x}}
\def\hy{\hat{y}}
\def\hz{\hat{z}}

\begin{document}

\title{Topological Entanglement Entropy of Fracton Stabilizer Codes}
\author{Han Ma}
\affiliation{Department of Physics, University of Colorado at Boulder, Boulder CO 80309, USA}
\author{A.T. Schmitz}
\affiliation{Department of Physics, University of Colorado at Boulder, Boulder CO 80309, USA}
\author{S. A. Parameswaran}
\affiliation{The Rudolf Peierls Centre for Theoretical Physics, University of Oxford, Oxford OX1 3NP, UK}
\affiliation{Department of Physics and Astronomy, University of California, Irvine, CA 92697, USA}
\author{Michael Hermele}
\affiliation{Department of Physics, University of Colorado at Boulder, Boulder CO 80309, USA}
\affiliation{Center for Theory of Quantum Matter, University of Colorado at Boulder, Boulder CO 80309, USA}
\author{Rahul M. Nandkishore}
\affiliation{Department of Physics, University of Colorado at Boulder, Boulder CO 80309, USA}
\affiliation{Center for Theory of Quantum Matter, University of Colorado at Boulder, Boulder CO 80309, USA}

\begin{abstract}
Entanglement entropy provides a powerful characterization of two-dimensional gapped topological phases of quantum matter, intimately tied to their description by topological quantum field theories (TQFTs). Fracton topological orders are three-dimensional gapped topologically ordered states of matter that lack a TQFT description. We show that three-dimensional fracton phases are nevertheless characterized, at least partially, by universal structure in the entanglement entropy of their ground state wave functions. We explicitly compute the entanglement entropy for two archetypal fracton models --- the `X-cube model' and `Haah's code' --- and demonstrate the existence of a non-local contribution that scales linearly in subsystem size. We show via Schrieffer-Wolff transformations that this piece of the entanglement entropy of fracton models is robust against arbitrary local perturbations of the Hamiltonian. Finally, we argue that these results may be extended to characterize localization-protected fracton topological order in {\it excited} states of disordered fracton models.
\end{abstract}
\maketitle

\section{Introduction}
The study of topological order is a major theme of modern condensed matter physics. Gapped topologically ordered states of matter are characterized by properties~\cite{wen1, wen2, wen3, wen4} such as ground state degeneracy on manifolds of non-trivial topology, the inability to distinguish distinct such ground states via local measurements, and the existence of local excitations that cannot be created by purely local operators, leading to fractionalization of statistics. In two spatial dimensions there exists a fairly complete understanding of topological order, described via topological quantum field theory (TQFT) and related ideas.~\cite{wenbook} However, our understanding of topological order in {\it three} spatial dimensions remains incomplete --- a lacuna brought into sharp relief by the development of the {\it fracton} models~\cite{Chamon2005,Bravyi2011,Haah2011,Yoshida2013,Vijay2015,Vijay2016}. These models exhibit the characteristic properties of topological order, but lack a description in terms of TQFTs. As such, they represent a new chapter in the story of topological order, and have begun to draw intensive interest~\cite{Williamson2016, Pretko1, Pretko2, Vijay2017, Hermele, Prem2017, Kim2017, Hsieh2017, Slagle2017, PremPretko}. 

In a parallel line of development, ideas from quantum information have been transplanted to the field of correlated systems, providing potent methods with which to characterize and study complex quantum many-body systems. Foremost among these new tools is the use of quantum entanglement -- specifically, its characterization in terms of {\it entanglement entropy} --- in developing the understanding of topological order (see e.g. Ref.~\onlinecite{GroverReview} for a review). For two-dimensional gapped phases, the entanglement entropy contains a universal subleading `constant' term~\cite{KitaevPreskill, LevinWen}. This is intimately related to the TQFT description of the topological phase and provides a partial characterization of the nature of topological order. It is thus natural to ask, what insights are afforded by studying entanglement in fracton phases of matter? A previous work\cite{Lu2017} has introduced certain bounds on the non-local part of the entanglement entropy in fracton models.

In this work, we discuss the entanglement entropy in fracton phases. 
We begin by reviewing some basic facts about entanglement entropy in lattice models, and the formalism developed to study the entanglement structure of stabilizer codes.
We then explicitly calculate the entanglement entropy for several $d=3$ models, including two paradigmatic fracton phases: the `X-cube model' and `Haah's code'.  Both fracton phases are found to exhibit a term of non-local origin in the entanglement entropy that scales linearly with subsystem size, with a coefficient that we calculate. We refer to this term as topological entanglement entropy. 

We also point out that the existence of topological entanglement entropy indicates that on the boundary of the subregion, certain non-local constraints act on the ground state wavefunction. We explicitly identify these constraints for the X cube model.  This provides an argument that the topological entanglement entropy is robust under arbitrary local perturbations of the Hamiltonian. Consider a perturbation $H'=\lambda V$,
where $\lambda$ is a control parameter, and $V$ is of unit norm and contains perturbations local in real space which do not commute with $H(\lambda=0)$. When $\lambda \neq 0$, the stabilizers are no longer eigenoperators of the groundstate.  However, for $\lambda$ sufficiently small that it does not induce a phase transition as increased from zero, the new ground state is related to the $\lambda = 0$ ground state by a unitary transformation
\begin{equation}
| \Psi(\lambda) \rangle = U | \Psi(\lambda = 0) \rangle \text{,} \label{eq:localU}
\end{equation}
where $U$ is a local unitary operator which can in principle be constructed by adapting the method of Schrieffer-Wolff transformations~\cite{BVL} (see Appendix~\ref{app:SW} for details). Non-local constraints on the wavefunction cannot be altered by local unitary transformations; thus the topological entanglement entropy (being of non-local origin) must be invariant under local perturbations. 

Finally we generalize this argument to excited states of disordered fracton models, which we argue can display localization protected fracton order, characterized by topological entanglement entropy.

\section{Review of topological entanglement entropy}
\label{sec:reviewtopo}

We begin by reviewing some essential facts about entanglement entropy, focusing in particular on topological contributions and how to extract them. We will consider a system in a topological phase, with $|\Psi\rangle$  one of its degenerate ground states. We divide the system into a small subregion $A$ and `everything else' ($B$), and construct the reduced density matrix $\rho_A = \textrm{Tr}_B |\Psi\rangle \langle \Psi |$. The entanglement entropy is then defined to be the von Neumann entropy of $\rho_A$,
\begin{equation}
S = - \textrm{Tr} \rho_A \log_2 \rho_A,
\end{equation}
(Here and throughout, we measure logarithms in base $2$ to remove unwieldy factors of $\ln 2$.)
In a gapped phase --- such as the fracton phases considered in this paper --- the entanglement entropy of a ground state is expected to follow an `area law' i.e. to be proportional at leading order to the surface area of the subregion $A$. If $A$ has linear size $R$, then we may expand in powers of $R$, viz.
\begin{equation}\label{eq:Sexp}
S = A_1 R^{d-1} + A_2 R^{d-2}+\dots
\end{equation}
where $d$ is the spatial dimension. The leading term is non-universal and dominated by short-distance physics. The `topological' information is contained in the subleading corrections; the challenge is to extract it.

For concreteness, let us review how this works in $d=2$, where we have
\begin{equation}\label{eq:Sexp2d}
S =  A_1 R  - c \gamma + \dots,
\end{equation}
and the  topological contribution is the constant piece, $S_{\text{topo}} = -c \gamma$, with $c$ the number of connected components of the boundary of $A$. In principle it appears as though extracting $\gamma$ should be a straightforward exercise: we should simply compute $S$, and extract its constant contribution and identify it as topological using the scaling form Eq. (\ref{eq:Sexp2d}). In practice, however, most systems of interest and all those we are concerned with in this paper are defined on a lattice. Then subregions often have sharp corners that can lead to non-universal constant contributions in Eq. (\ref{eq:Sexp2d}), assuming the entanglement cut is a sequence of edges forming a continuous path. 
This complicates a direct identification of $S_{\textrm{topo}}$ from the scaling of the bipartite entanglement entropy in Eq. (\ref{eq:Sexp2d}).

\begin{figure}[t]
\includegraphics[width=0.9\columnwidth]{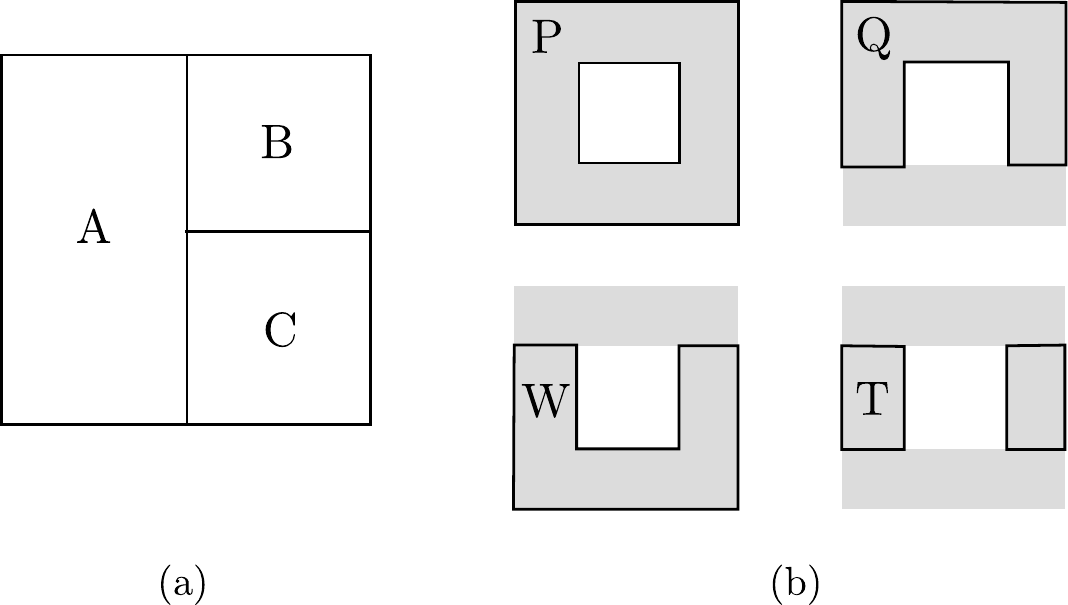}
\caption{Illustration of the two types of prescriptions used to obtain the topological entanglement entropy.  An ABC prescription is illustrated in (a), while (b) illustrates a PQWT prescription.  \label{fig:ABCD}}
\end{figure}

However, there are prescriptions~\cite{KitaevPreskill, LevinWen} to extract topological entanglement by suitably combining the results for a variety of bipartitions. Two such prescriptions in two dimensions are illustrated in Fig.~\ref{fig:ABCD}.  We refer to the type of prescription illustrated in Fig.~\ref{fig:ABCD}(a) as an  ABC prescription, where $A$, $B$ and $C$ are three disjoint regions, and the topological entanglement entropy is given by
\begin{equation}
S^{ABC}_{{\rm topo}} = S_A + S_B + S_C - S_{AB} - S_{BC} - S_{AC} + S_{ABC} \text{,} \label{eq:stopo-abc}
\end{equation}
where $AB \equiv A \cup B$ and so on.
An alternate prescription is shown in Fig.~\ref{fig:ABCD}(b) and referred to as a PQWT prescription.  In this case, the regions $P$, $Q$, $W$ and $T$ have the properties that $P = Q \cup W$ and $T = Q \cap W$, and the topological entanglement entropy is given by
\begin{equation}
S^{PQWT}_{{\rm topo}} = S_P - S_Q - S_W + S_T \text{.} \label{eq:stopo-pqwt} 
\end{equation}
In both cases, these linear combinations of entropies are chosen to ensure that the dependence 
on local contributions from boundaries, including corner contributions, cancels out.  These topological entanglement entropies are related to Eq.~(\ref{eq:Sexp2d}) by $S^{ABC}_{{\rm topo}} = - \gamma$ and $S^{PQWT}_{{\rm topo}} = -2 \gamma$, which can be understood by counting the number of connected components in the boundaries of the various regions involved.

Both ABC type and PQWT type prescriptions have been generalized ~\cite{CastelnovoChamon2008, GroverTurner} to $d=3$, and we will make use of these generalizations in this paper.  While different geometries and topologies of the regions are possible in these generalizations, in ABC type prescriptions we require the regions $A$, $B$ and $C$ be disjoint.  In contrast, in the PQWT type prescriptions we employ, the set theoretic properties $P = Q \cup W$ and $T = Q \cap W$ will always be satisfied.

Much progress has been made in understanding topological order and topological entanglement in $d=2$ by linking these ideas to TQFTs. For instance, we may understand the dependence of the topological contribution on the number of connected components of the boundary by recognizing that $S_{\text{topo}}$ reflects the additional information obtained by counting field lines of an gauge field constrained by a lattice analog of Gauss's law. Entanglement entropy in turn can provide an important tool for extracting TQFT data~\cite{GroverReview}, such as the braiding and statistics of the fractionalized excitations~\cite{BraidingEnt}.

Far less is known about three-dimensional topological phases, particularly in situations where there is no TQFT description. In this paper, we study the topological entanglement entropy of fracton models that are paradigmatic examples of these unexplored topological orders. To do so, we leverage an approach complementary to TQFT, namely we study zero-correlation-length Hamiltonians that are sums of commuting projectors, which describe special exactly-solvable points within some topological phases.  More specifically, we consider `stabilizer Hamiltonians,' where the commuting terms in the Hamiltonian are products of Pauli operators.
Our  strategy will be to first compute entanglement entropy at solvable points, and then argue that the entanglement entropy has a topological contribution that persists under perturbation, using various three-dimensional generalizations of the ABC and PQWT prescriptions for obtaining topological entanglement entropy. 

\section{Topological entanglement entropy in stabilizer codes}

\subsection{Entanglement entropy for stabilizer codes}\label{sec:ent}

We  begin our analysis by reviewing the method developed in Ref.~\onlinecite{zou-haah,hamma2005bipartite,fattal2004entanglement} for calculating the entanglement entropy for the ground states of stabilizer codes, and discussing the straightforward extension to arbitrary eigenstates.
This extension will be used in Sec.~\ref{sec:LPQO} when we discuss localization protected fracton topological order in excited states. We will illustrate the use of the stabilizer formalism on the toric code before applying these techniques to fracton models. Readers familiar with the stabilizer formalism for ground-state entanglement computations may wish to skip ahead to Sec.~\ref{sec:x-cube}.

Throughout, we will consider systems of spin-$1/2$ degrees of freedom (`qubits') that reside on either vertices or links of $d$-dimensional hypercubic lattices. The Hamiltonians is built solely of terms containing either  $X$ type stabilizer operators that are products of Pauli matrices $X_i$, or $Z$ type stabilizer operators that are products of Pauli matrices $Z_i$. The stabilizer operators all mutually commute, so that the Hamiltonian is exactly solvable. By construction, both $X$- and $Z$-type stabilizers square to the identity and hence have eigenvalues $\pm 1$.

Let us take $\mathcal{S} = \{O_s\}$ to be a set of mutually commuting stabilizers. Then, a state $\ket{\psi}$ is stabilized by $\mathcal S$ if it satisfies
\begin{equation}
O_s \ket{\psi} = \ket{\psi},
\end{equation}
for all $O_s \in \mathcal{S}$. 
The set of states stabilized by $\mathcal{S}$ is called a stabilizer code and is the ground state manifold of the Hamiltonian, $H_{\text{stab}} = -\sum_{s} J_s O_s$ with $J_s>0$. Here, we are assuming that the $O_s$ operators are local, but, below, we relax this assumption.

It is useful to  consider the Abelian group $G$ that is multiplicatively generated by the stabilizers in $\mathcal{S}$. We assume that all the stabilizers in $\mathcal{S}$ are independent and collectively act on exactly $N$ spins-$1/2$. For open boundary conditions, this is generally expected to be the case for a maximal set of independent local stabilizers.  For periodic boundary conditions, we usually need to include some non-local stabilizers that, for instance, wrap non-contractible loops. Elements $g\in G$ may be labeled by a binary vector $\vec{n} = \{n_1, n_2,\dots, n_{| \mathcal{ S}|}\}$, with $n_i \in\{0,1\}$ via
\begin{equation}\label{eq:gofndef}
g(\vec{n}) =  O_1^{n_1} O_2^{n_2}\ldots O_{|\mathcal{S}|}^{n_{|\mathcal{ S}|}},
\end{equation}
where $|\mathcal{S}|$ is the number of stabilizers in $\mathcal{S}$. Note that $\mathcal{S}$ is a generating set. Because $G$ is Abelian, we may label states in the Hilbert space by their eigenvalues of the group elements $g\in G$.

We note that $G$ can be viewed as a vector space over the two-element field ${\mathbb F}_2$, a fact that will be useful in our approach to numerical calculation of entanglement entropy.  This statement holds for any group, like $G$, that is isomorphic to a product of ${\mathbb Z}_2$ factors.  Vector addition corresponds to multiplication of stabilizers.  The zero vector corresponds to the group identity.  Scalar multiplication is trivial; multiplication by $1 \in {\mathbb F}_2$ is the identity operation, and multiplication by $0 \in {\mathbb F}_2$ sends any element of $g \in G$ to the identity of $G$.  We will use both vector space and group language to describe operations in $G$, and will do this without comment when the meaning is clear from the context.  We note that, in vector space language, the set ${\mathcal S}$ is a basis.

It remains to determine the size of the group $G$, that we denote $|G|$. We further suppose that there is a {\it unique} eigenstate $\ket{\psi}$ of eigenvalue $+1$ for all $g\in G$; if this is not the case, then we may add elements to $\mathcal{S}$ until this is so. Then, $G$ cannot contain any pure scalar element $\eta\neq 1$ since such an element must have an eigenvalue that is not $1$ (a scalar element is an element proportional to the identity). We may then write the projector onto $\ket{\psi}$ as
\begin{equation}\label{eq:projectorgs}
|\psi \rangle \langle \psi |  = \frac{1}{|G|} \sum_{\vec{n}} g(\vec{n}).
\end{equation}
(To see this, observe using group properties that the RHS squares to itself, i.e. is a projector, and acts as the identity on $\ket{\psi}$, whence the result Eq. (\ref{eq:projectorgs}) follows by the uniqueness of the ground state.)
Taking the trace on both sides, we find
\begin{equation}
1 = \frac{1}{|G|} \sum_{\vec{n}} \textrm{Tr} g(\vec{n}) =    \frac{1}{|G|} \textrm{Tr} I = \frac{2^N}{|G|}, 
\end{equation}
where we use the uniqueness of the identity $I\in G$ and the fact that any non-identity element in $G$ is traceless; the final step follows simply from the fact that the Hilbert space is the tensor product of $N$ spins-$1/2$. Thus, we see that the size of $G$ is the full dimension of the Hilbert space, $|G| = 2^N$.

From this and Eq. \eqref{eq:gofndef} we conclude that there are $N$ independent stabilizers in $\mathcal{S}$. i.e. $|\mathcal{ S}| = N$. Furthermore, we see that the group $G$ must be isomorphic to  $\mathbb{Z}_2^{\otimes N}$, the group of spin-flips on $N$ spins (intuitively, we may think of the `spin' as the eigenvalue of the stabilizer $\mathcal{O}_s$.) Therefore, we may completely label eigenstates of $H_{\text{stab}}$ in terms of irreducible representations of the spin-flip group~\cite{hamma2005bipartite}. The irreps are one-dimensional and are labeled by a binary string $\vec{k}$ of length $N$, and defined by the map $\rho_{\vec{k}}[g(\vec{n})] = (-1)^{\vec{k}\cdot\vec{n}}$; in other words, the eigenvalue of a group element $g$ in state $\vec{k}$ is given by  $(-1)^{\vec{k}\cdot\vec{n}}$. It then follows from standard orthogonality relations that the density matrix of state $\ket{\vec{k}}$ can be written as
\begin{equation} \label{eq:proj}
|\vec{k} \rangle \langle\vec{k} | = \mathcal{P}^{(\vec{k})} = \frac{1}{|G|} \sum_{\vec{n}} (-1)^{\vec{k}\cdot\vec{n}} g(\vec{n}).
\end{equation}
In this notation, the ground state $\ket{\psi}$ corresponds to the vector $\vec{k} = (0,0,\ldots)$.

We will now demonstrate how to compute the entanglement entropy of a subregion for {\it any} such eigenstate $\ket{\vec{k}}$ of $H_{\text{stab}}$. Consider any bipartition $(A,B)$ as described above. Then, we see that the reduced density matrix of subregion $A$ in state $\ket{\vec{k}}$ is 
\begin{equation}\label{eq:rhoastep1}
\rho_A = \textrm{Tr}_B |\vec{k} \rangle \langle \vec{k}|=\frac{1}{|G|} \sum_{\vec{n}} (-1)^{\vec{k}\cdot\vec{n}} \textrm{Tr}_B g(\vec{n}). 
\end{equation}
Any element $g$ that is {\it not} equal to the identity on $B$ must contain at least one $X$ or $Z$ operator acting in $B$ and consequently will have vanishing trace in $B$. Thus, the only non-zero contributions to the sum in Eq. \eqref{eq:rhoastep1} are from operators supported only on $A$, i.e. equal to the identity on $B$. Operators in the sum supported only on $A$ form a subgroup $G_A$, with elements $g(\vec{n}_A)$, and irreps labeled by $\vec{k}_A$, where we have made the obvious generalizations of notation. Since the identity on $B$ has trace $\textrm{Tr} I_B = 2^{N_B}$ where $N_B$ is the number of spins in $B$, we have (using $|G| = 2^N = 2^{N_A+N_B}$) that
\begin{equation}\label{eq:rhoastep2}
\rho_A= \frac{2^{N_B}}{|G|} \sum_{n_A}  (-1)^{\vec{k}_A \cdot \vec{n}_A} g(\vec{n}_A) =\frac{ |G_A| }{2^{N_A}} \mathcal{P}^{\left(\vec{k}_A\right)}_A.
\end{equation}
Note that the group $G_A$ only includes {\it complete} stabilizers in $A$, since stabilizers that `dangle' across the entanglement cut are not equal to the identity on $B$. Since Eq. \eqref{eq:rhoastep2} expresses $\rho_A$ as a projector, its entanglement entropy follows straightforwardly:
\begin{equation}\label{eq:stabEEform}
S_A = N_A - \log_2 |G_A|.
\end{equation}
Thus, the computation of entanglement entropy of a subregion $A$ reduces to that of determining the size of the subgroup $G_A \subset G$ that consists of stabilizers contained entirely within $A$.  In vector space language, $\log_2 | G_A| = \dim G_A$, so
\begin{equation}\label{eq:stabEEform_vs}
S_A = N_A - \dim G_A \text{.}
\end{equation}

Viewed as an ${\mathbb F}_2$ vector space, $G_A = G^Z_A \oplus G^X_A$, where $G^Z_A$ and $G^X_A$ are vector spaces of $Z$ and $X$ stabilizers, respectively.  Therefore 
\begin{equation}
\dim G_A = \dim G^Z_A + \dim G^X_A \text{,}
\end{equation}
so that $Z$ and $X$ stabilizers can be treated separately in computing $S_A$.

\begin{figure}[t]
\includegraphics[width=.25\textwidth]{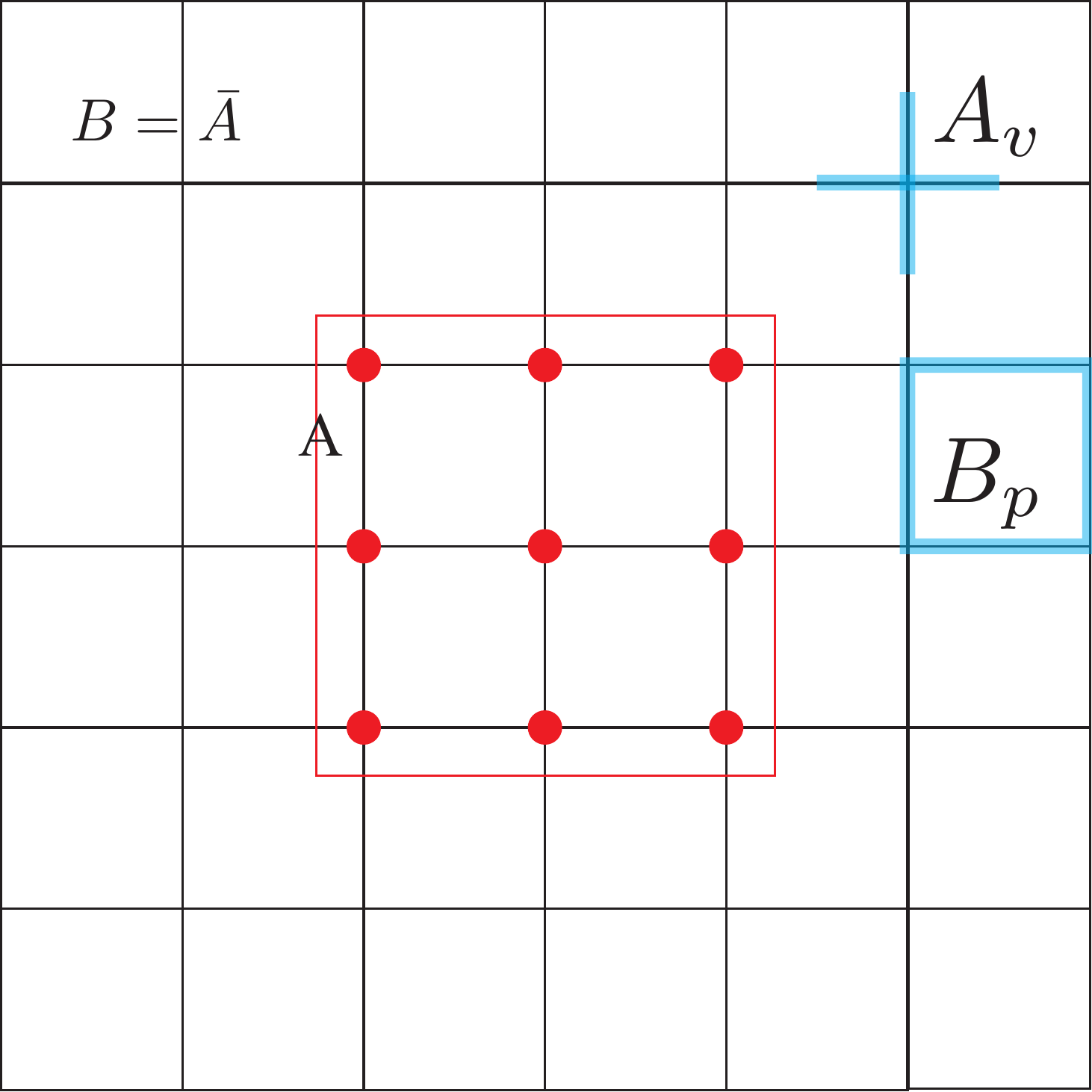}
\caption{Two-dimensional toric code on the square lattice.  The region enclosed by red the line is the subregion $A$, which has size $2 \times 2$ measured in edges of the lattice. Spins on links cut by the red line lie outside $A$.  The vertex and plaquette terms $A_p$ and $B_v$ are also shown. \label{fig:tc_entropy}}
\end{figure}

\subsection{Local and non-local stabilizers}
\label{sec:loc-nonloc}

Here, we briefly discuss some further properties of the stabilizer group $G_A$ in connection with entanglement entropy and topological entanglement entropy.  These properties are then used in the numerical procedure for computing entanglement entropy described in Sec.~\ref{sec:calculating}.  Moreover, we show that $S^{PQWT}_{{\rm topo}}$ is entirely determined by counting non-local stabilizers, while in many cases $S^{ABC}_{{\rm topo}}$ can be determined by counting local stabilizers.

Given a stabilizer Hamiltonian, we obtain a set ${\cal S}^Z_{loc}$ of local $Z$ stabilizers, which are just the terms of the Hamiltonian.  (For simplicity of discussion we focus for the moment on $Z$ stabilizers; identical statements hold for $X$ stabilizers.)  Local $Z$-stabilizers supported entirely in $A$ generate a subgroup $G^Z_{A,loc} \subset G^Z_A$. If $G^Z_A = G^Z_{A, loc}$, we say that $G^Z_A$ is \emph{locally generated}.   In this case,  $\dim G^Z_A$ can be obtained by simple counting, accounting for possible constraint equations satisfied by the local stabilizers.
{(A constraint is a product of local stabilizers which evaluates to the identity, i.e. $\prod_{O_s \in F} O_s = I$ for some subset $F \subseteq \mathcal S$.)}
In general,  $G^Z_{A,loc} \neq G^Z_A$. If $g \in G^Z_A$ but $g \notin G^Z_{A, loc}$, we call $g$ a \emph{non-local stabilizer}.  In addition to $\dim G^Z_A$, we will be interested in the number of independent non-local stabilizers, which is defined by
\begin{equation}
\Omega^Z_A \equiv \dim G^Z_A/G^Z_{A,loc} \text{.}
\end{equation}
By taking the quotient, we are counting non-local stabilizers up to multiplication by local stabilizers. That is, two non-local stabilizers in $G^Z_A$ that are related by a product of local stabilizers in $G^Z_{A,loc}$ are not considered independent in this counting. We have
\begin{equation}
 \dim G^Z_A = \dim G^Z_{A,loc} + \Omega^Z_A  \text{.} \label{eqn:dim_omega_reln}
 \end{equation}

Now we consider topological entanglement entropy $S^{ABC}_{{\rm topo}}$, obtained in an ABC type prescription as discussed in Sec.~\ref{sec:reviewtopo}.  We assume that the stabilizer groups $G_A$, $G_B$, $G_C$, $G_{AB}$, and so on, are all locally generated.  Moreover, we assume the generators of these groups are drawn from a set of local stabilizers that do not obey any local constraints.  This holds trivially for the $d=2$ toric code and Haah's code, where the stabilizers do not obey any local constraints.  For the X-cube model, the vertex stabilizers do obey local constraints, but for suitable regions $A$, $B$, $C$ it is possible to use only $xy$ plane and $xz$ plane vertex stabilizers, which do not obey local constraints, as shown in Appendix~\ref{app:xcube-locgen}.  Under these assumptions, $-S_{{\rm topo}}$ is simply the number of independent local stabilizers that have non-trivial support in each of the disjoint regions $A$, $B$ and $C$.  The contributions of other local stabilizers to $S^{ABC}_{{\rm topo}}$ cancel out.  For instance, suppose some local stabilizer is supported entirely in $A$.  Then it is also contained in $AB$, $AC$ and $ABC$, and it contributes $+1$ to each of $\dim G_A$, $\dim G_{AB}$, $\dim G_{AC}$ and $\dim G_{ABC}$.  These contributions cancel in $S_{{\rm topo}}$.  Similarly, if a local stabilizer has non-trivial support in both $A$ and $B$, but not $C$, then it is contained in both $AB$ and $ABC$, and its contribution to $S_{{\rm topo}}$ cancels.  A more careful argument for this result is given in Appendix~\ref{app:nll}.

In contrast, $S^{PQWT}_{{\rm topo}}$, the topological entanglement entropy  obtained via a PQWT type prescription is determined in many cases entirely by non-local stabilizers. Here we give a rough argument that all contributions of local stabilizers cancel out; a more complete treatment is given in Appendix~\ref{app:nll}. First recall 
 that $P = Q \cup W = Q \cup W \cup T$, so it is enough to consider different types of local stabilizers supported entirely on $P$.  Suppose a local stabilizer is supported entirely in $T$, then it is contained in all four regions, and its contribution to $S^{PQWT}_{{\rm topo}}$ cancels.  Now suppose a stabilizer is contained in $Q$ but not in $T$, then it is also contained in $P$ but not in $W$, so its contribution cancels.  This covers all the possibilities for local stabilizers.  Because $N_P + N_T = N_Q + N_W$,
we have the result
 \begin{equation}
 S^{PQWT}_{{\rm topo}} = - \Omega_P + \Omega_Q + \Omega_W - \Omega_T + \Delta_{PQWT} \text{,} \label{eqn:spqwt-nonlocal}
 \end{equation}
 where $\Omega_P = \Omega^Z_P + \Omega^X_P$ is the total number of non-local stabilizers in $P$, and similarly for the other regions. $\Delta_{PQWT}$ is a correction associated with non-local constraints among local stabilizers.  In Appendix~\ref{app:nll}, we derive Eq.~(\ref{eqn:spqwt-nonlocal}), and show that $\Delta_{PQWT} = 0$ for all the models discussed in this paper except the $d=3$ toric code.

\subsection{Calculating the entanglement entropy}
\label{sec:calculating}

In many cases, it is possible to determine $|G_A|$ and hence the entanglement entropy $S_A$ analytically.  However,  this is not always straightforward, and numerical calculation is useful as a check on other methods and sometimes as a primary means of determining the entanglement entropy.  Here, we describe a numerical procedure to determine $\dim G_A = \log_2 |G_A|$.  We are always interested in the case where $A$ is a subset of a thermodynamically large region. Under a suitable assumption discussed below, which can be verified for particular models of interest, $\dim G_A$ does not depend on global properties of the large region containing $A$.
 
 To proceed, we choose a finite enclosing region $B$ with $A \subset B$.  It is obvious that $G^Z_A \subset G^Z_B$.  We make the assumption that it is always possible to choose $B$ so that $G^Z_B$ is locally generated.  This assumption implies $G^Z_A \subset G^Z_{B, loc}$; that is, all stabilizers in $A$ are products of local stabilizers in $B$.  We show in Appendix~\ref{app:enclosing} that this assumption holds for the stabilizer codes studied in this paper.

Now we introduce $F^Z_{B, loc}$, the group of formal products of local $Z$ stabilizers supported on B.  We let $M$ be the number of local $Z$ stabilizers supported entirely in $B$, and denote these operators by $O_s$ ($s = 1,\dots, M$).  A general product of these stabilizers is $(O_1 )^{n_1} \dots (O_M )^{n_M}$, where $n_i = 0, 1$. There are $2^M$ such products, and in the group $F^Z_{B,loc}$, we treat them all as distinct elements, so that $F^Z_{B, loc} \simeq {\mathbb Z}_2^{\otimes M}$. In general, two formal products in $F^Z_{B, loc}$ can correspond to the same operator, because there can be constraints among the local stabilizers.   
There is a linear map
\begin{equation}
\phi_Z: F^Z_{B, loc} \rightarrow P^Z_B \text{,}
\end{equation}
where $P^Z_B$ is the group (or $\mathbb{F}_2$ vector space) generated by all $Z$ Pauli operators supported on $B$. The map $\phi_Z$ is defined by replacing the $O_s$ in a formal product with their expressions in terms of Pauli operators.  $G^Z_{B,loc}$ is a subspace of $P_B^Z$, and moreover $G^Z_{B,loc} = \textrm{Im} \phi_Z$.  
If $f \in \textrm{Ker} \phi_Z$, then $f$ is a formal product of stabilizers that evaluates to the identity operator. This happens when the stabilizers obey some constraint equations. Indeed, the number of independent such constraints is precisely $\dim \textrm{Ker}\phi_Z$.

We note that $P_B^Z = P^Z_{B-A} \oplus P_A^Z$, where $B - A$ is the complement of $A$ in $B$.  
We thus have the projection map
\begin{equation}
\pi_{B-A} : P_B^Z \rightarrow P^Z_{B-A},
\end{equation}
defined by
\begin{eqnarray}
\pi_{B-A}(g) = \left\{ \begin{array}{ll}
1 & g \in P_A^Z \\
g & g \in P_{B-A}^Z \end{array}
 \right. \text{.}
\end{eqnarray}
Elements in the kernel of $\pi_{B-A}$ are products of $Z$ Pauli operators  supported entirely on $A$.

The last step is to consider the composition $\pi_{B-A} \circ \phi_Z$. Suppose $f \in \operatorname{Ker} (\pi_{B-A} \circ \phi_Z)$, but $f \notin \operatorname{Ker} \phi_Z$. This means that $f$ corresponds to a non-trivial element of $G^Z_A$. We are finally able to express the number of stabilizers in $A$ as\footnote{To prove this result, we let $\{a_1, \dots, a_n \}$ be a basis for $\operatorname{Ker} \phi_Z$ and extend it to a basis $\{ a_1, \dots, a_n, b_1, \dots, b_m \}$ for $\operatorname{Ker} \pi_{B-A} \circ \phi_Z$.  It can then be checked that $\{ \phi_Z(b_1), \dots, \phi_Z(b_m) \}$ is a basis for $G^Z_A$, and the result follows}
\begin{equation}
\dim G^Z_A = \dim \operatorname{Ker} \pi_{B-A} \circ \phi_Z - \dim \operatorname{Ker} \phi_Z \text{.} 
\end{equation}
Here, the last term is subtracted to avoid incorrectly counting constraints among local stabilizers as non-trivial elements of $G^Z_A$.  The linear maps $\phi_Z$ and $\pi_{B-A}$ can be constructed explicitly as matrices, and numerical linear algebra methods can then be used to compute the dimensions of the kernels.  Our calculations were done using routines for linear algebra over ${\mathbb F}_2$ in Mathematica.

There is a minor modification of the above approach that significantly reduces the computational effort required. 
We consider the subspace $F^Z_{A,loc} \subset F^Z_{B,loc}$, which consists of formal products of stabilizers supported entirely in $A$.  Then, we have $F^Z_{B,loc} = F^Z_{A,loc} \oplus \mathcal{F}$ where $\mathcal{F}$ consists of formal products of stabilizers in $B$ that either lie completely outside $A$ or are not fully contained in $A$. Taking $f \in {\mathcal F}$, if $f \in \operatorname{Ker} \pi_{B-A} \circ \phi_Z$ with $f \notin \operatorname{Ker} \phi_Z$, then $\phi_Z(f)$ is a non-trivial stabilizer in $G^Z_A$.  The number of such stabilizers is
\begin{equation}
\tilde{\Omega}^Z_A =\dim\operatorname{Ker} \pi_{B-A} \circ \phi_Z|_\mathcal{F} -\dim\operatorname{Ker} {\phi_Z}|_{\mathcal{F}} \text{.}
\end{equation}
It can happen that $\phi_Z(f)$ is a local stabilizer in $G^Z_A$; this can happen when the local stabilizers obey some  local constraints, so that some local stabilizers in $G^Z_A$ can be written as a product of local stabilizers not supported entirely in $A$.  We let $K^Z_{A,loc} \subset G^Z_{A,loc}$ be the subspace generated by such local stabilizers in $A$. If the local stabilizers generating $G^Z_{B,loc}$ obey no local constraints, then $\tilde{\Omega}^Z_A = \Omega^Z_A$.  More generally,
\begin{equation}
\Omega^Z_A = \tilde{\Omega}^Z_A - \dim K^Z_{A, loc}  \text{,}  \label{eq:omega_A^Z}
\end{equation}
which determines $\dim G^Z_A$ via Eq.~(\ref{eqn:dim_omega_reln}).

We use this method to compute topological entanglement entropy for Haah's code as discussed in Sec.~\ref{sec:haah}.  In addition, we employ the same method to check our results for the X-cube model in Sec.~\ref{sec:x-cube}.

\subsection{Simple example: topological entanglement entropy of $d=2$ toric code}
\label{sec:toric2d}

To illustrate the use of the stabilizer formalism, we now use it to compute the entanglement entropy of the $d=2$ toric code model~\cite{kitaev2003fault}. 
One qubit resides on each link of the square lattice, and the Hamiltonian is
\begin{equation}\label{eq:TCHam}
H_{\textrm{TC}} = - \sum_{v}A_v - \sum_{p} B_p,
\end{equation}
where $B_p$ is the product of the four $Z$ operators surrounding the plaquette $p$, and $A_v$ is the product of the four $X$ operators connected to vertex $v$.

We compute the entanglement entropy of a subsystem $A$ of size $R \times R$, shown in Fig.~\ref{fig:tc_entropy} for $R=2$. This region contains $N_A = 2(R+1)R$ spins. Additionally, there are $R^2$ plaquette terms and $(R-1)^2$ vertex terms confined entirely within subregion $A$; these are local stabilizers and they are all independent.  Using the fact that $A$ is simply connected, it can be seen easily that $G_A$ is generated by the plaquette and vertex stabilizers  supported entirely in $A$.  Moreover, these stabilizers obey no local constraints.  Therefore, $|G_{A}|= 2^{R^2 + (R-1)^2}$.

Using Eq. \eqref{eq:stabEEform}, we find that the entanglement entropy is
\begin{equation}
S_A =2(R+1)R-R^2- (R-1)^2= 4R-1 \text{.}
\end{equation}
Since the boundary has length $|\partial A| = 4R$ measured in lattice edges, we have $S_A = |\partial A| +S_{\text{topo}} $, where $S_{\text{topo}}=-1$ is the well-known topological entanglement entropy of the two dimensional toric code. 
This is something of an accident; this region has sharp corners, but the corner contributions happen to vanish.  To actually obtain $S_{\text{topo}}$, we can break region $A$ into three subregions $A$, $B$, $C$ as in Fig.~\ref{fig:ABCD}a, and apply the ABC prescription to obtain $S^{ABC}_{{\rm topo}}$ as in Eq.~(\ref{eq:stopo-abc}).

To illustrate features not arising in the above simple example, we calculate $S_P$  for an annulus-shaped region $P$ shown in Fig. (\ref{fig:ABCD}b).  There, due to the non-trivial topology, $G_P$ is no longer generated by local stabilizers supported entirely on $P$.  Using the terminology of Sec.~\ref{sec:loc-nonloc}, $G_P$ is generated by the local stabilizers in $P$, together with two non-local stabilizers.  These are products of plaquette and vertex stabilizers, respectively, over the hole in the annulus.   Taking the linear size of the hole to be $R$ and that of the exterior edges to be $3 R$, we obtain $S_{P} = 16R-6 = |\partial P| -4-2$, where the constant $-4$ is non-universal and contributed by vertex terms at four concave corners of the subsystem while the  $-2$ results from the number of non-local plaquette and vertex stabilizers acting on this subsystem. Although there are non-universal parts in this entropy due to the detailed geometry, the topological part of the entropy can be extracted by canceling all of those boundary contributions out via the PQWT prescription in Eq. (\ref{eq:stopo-pqwt}).  
 
A different perspective on the topological entanglement entropy, that further clarifies its robustness, is afforded by an understanding of the ground state wave function of the toric code as a condensate of closed loops of $\mathbb{Z}_2$ `electric' field lines. Each connected component of the entanglement surface intersects any loop an {\it even} number of times; this topological fact provides exactly one bit of extra information about the ground state, thereby reducing the entanglement entropy by a universal correction of $-1$ for each connected component of the entanglement surface. Similar ideas can be used to clarify the topological entanglement entropy of other topological orders in two dimensions by constructing their ground states as string-net condensates~\cite{LevinStringnet,LevinWen}.

\section{Topological Entanglement Entropy in $d=3$ }

As we have discussed, extracting the topological contribution to the entanglement entropy generally requires employing a prescription designed to cancel local contributions.  This becomes even more important in $d=3$, where Ref.~\onlinecite{GroverTurner} argued that even for a subregion $A$ with a smooth boundary, in the absence of parity and continuous rotation symmetry, every term in the expansion of $S_A$ in powers of the inverse linear size $R^{-1}$ acquires a local, non-universal contribution.  Continuous rotation symmetry is absent in fracton models, and, more seriously, the dynamics of fracton excitations leads to an expectation that fracton topological orders lack a continuum description with continuous rotation symmetry.  Discrete symmetries like parity may or may not be present in a given solvable fracton model, but certainly need not be present upon perturbing such a model to make it generic.

 \begin{figure}[t]
\includegraphics[width=.3\textwidth]{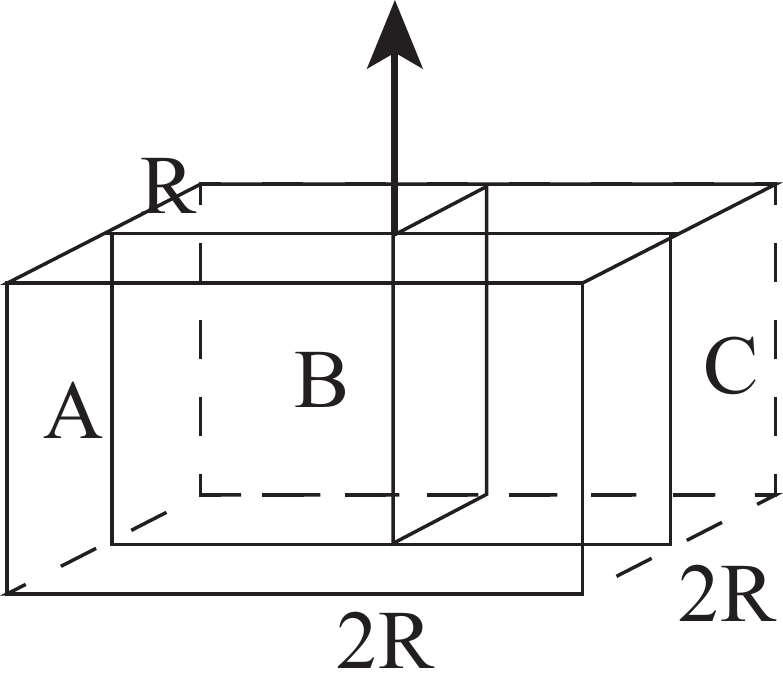}
\caption{Regions for a $d=3$ ABC prescription to compute topological entanglement entropy.  This choice of regions picks out a preferred axis (arrow).  $A$ is a  $2R \times R \times R$ rectangular prism, and regions $B$ and $C$ both have dimensions $R \times R \times R$.  The union $ABC$ of the three regions is a $2R \times 2R \times R$ rectangular prism, with $R$ the linear size along the preferred axis. \label{fig:ABCD_3d} }
\end{figure}

\begin{figure}[t]
\includegraphics[width=.4\textwidth]{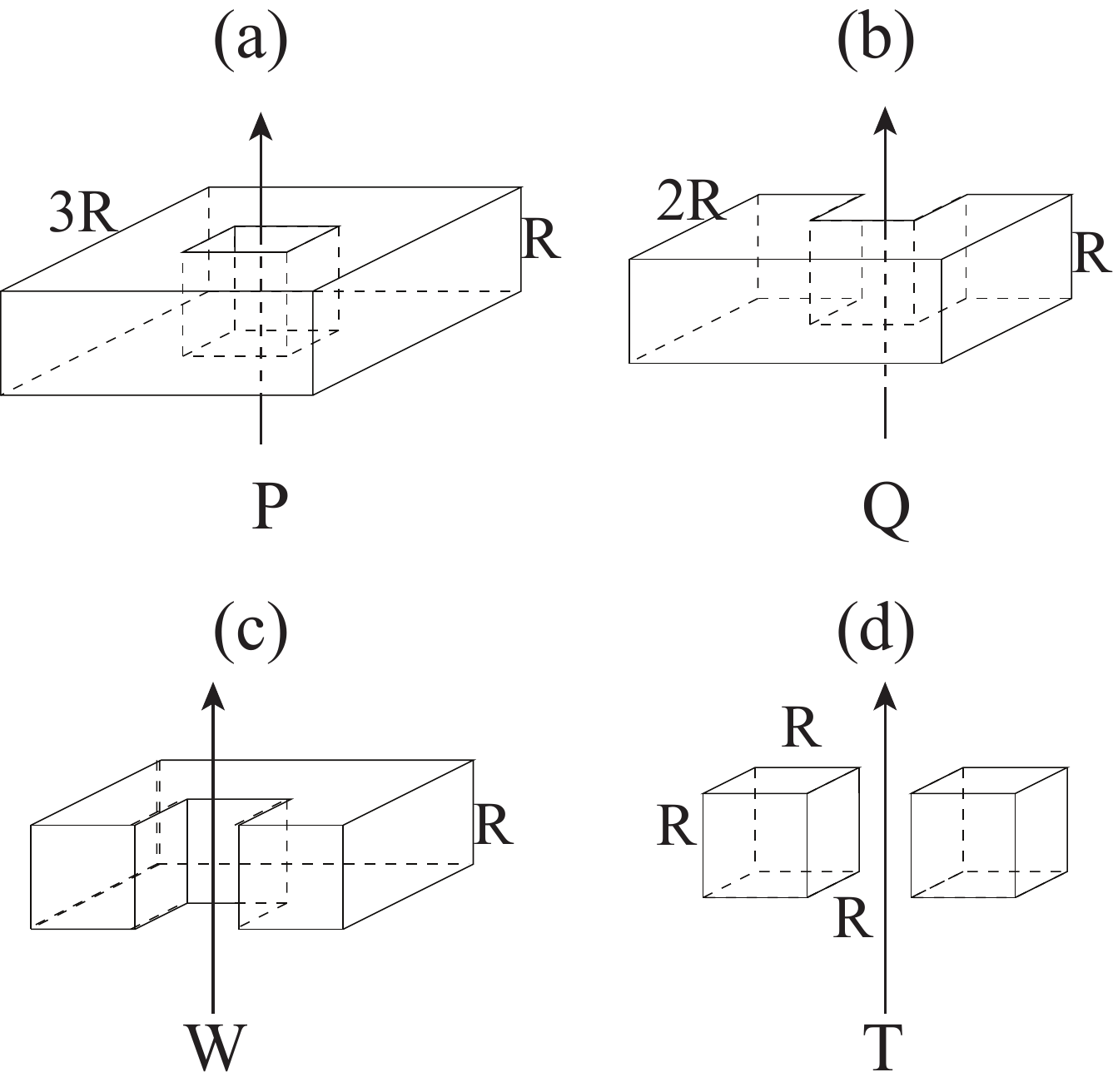}
\caption{\label{fig:ABCD_3d_2} Regions for a PQWT prescription with a preferred axis denoted by arrows in $d=3$ to extract topological entanglement entropy.  Lengths are measured in terms of the lattice distance (\emph{i.e.} the number of links). 
}
\end{figure}

Therefore, we rely on $d=3$ generalizations of the ABC and PQWT prescriptions discussed in Sec.~\ref{sec:reviewtopo}.  We use two different PQWT type prescriptions~\cite{CastelnovoChamon2008}, one is illustrated in Fig.~\ref{fig:ABCD_3d_2}, the other in Fig.~\ref{fig:ABCDHaah}.  A na\"{\i}ve extension of the $d=2$ ABC prescription (Fig.~\ref{fig:ABCD}(a)) is shown in Fig.~\ref{fig:ABCD_3d}.  As noted in Ref.~\onlinecite{GroverTurner}, this prescription fails to cancel local contributions from the two points where regions $A$, $B$ and $C$ all meet at the top and bottom boundaries.  This implies that $S^{ABC}_{{\rm topo}}$ is contaminated by non-universal contributions that are constant in $R$.  However, we will still employ this prescription, because in fracton models we will find a contribution to $S_{{\rm topo}}$ proportional to $R$, which is unaffected by the uncanceled constant local contributions.  While we do not use them in this paper, we note that Ref.~\onlinecite{GroverTurner} introduced different $d=3$ ABC prescriptions that do not suffer from this issue.

\begin{figure}[t]
\includegraphics[width=.3\textwidth]{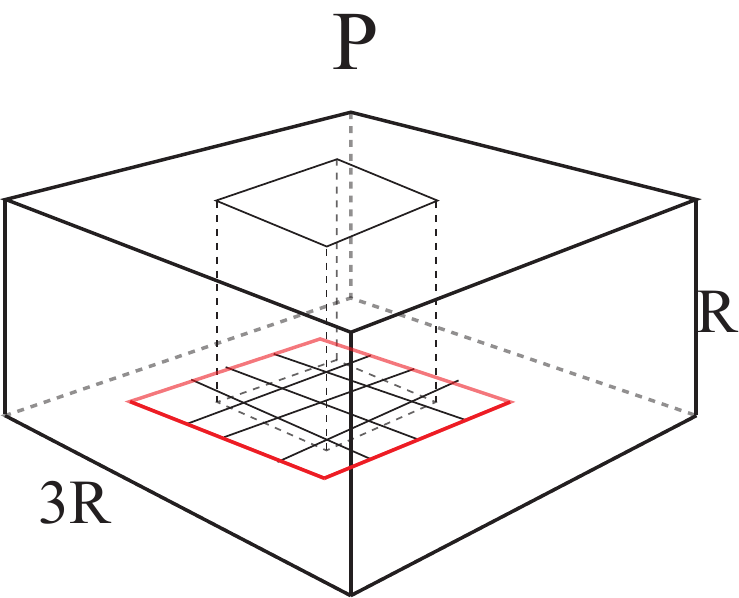}
\caption{Nonlocal stabilizer (red square) for $d=3$ toric code, which is a product of plaquette terms in the $B = \bar{P}$ subregion and those on the boundary but only acts nontrivially on the $P$ subsystem. \label{fig:operator_tc} }
\end{figure}

To illustrate our approach in $d=3$, we briefly discuss the calculation of topological entanglement entropy for the $d=3$ toric code, using the PQWT prescription illustrated in Fig.~\ref{fig:ABCD_3d_2}.  The Hamiltonian is given by Eq.~\eqref{eq:TCHam} but on the cubic lattice, so that the vertex terms involve six spins, and there are plaquette terms for each face of a cubic unit cell. As found in Eq.~\ref{eqn:spqwt-nonlocal}, $S^{PQWT}_{{\rm topo}}$ is given entirely by counting non-local stabilizers in each region.  Region $P$ has a single non-local stabilizer, which can be constructed by taking a product of local plaquette stabilizers over an $xy$ plane surface that cuts through the inner cube (\emph{i.e.} the ``hole'' at the center of $P$), as illustrated in Fig.~\ref{fig:operator_tc}.  Regions $Q$, $W$ and $T$ have no non-local stabilizers, so $S^{PQWT}_{{\rm topo}} = - \Omega_P = -1$. This is consistent with the fact that the ground state wave function is a loop condensate, with loops cutting the entanglement surface an even number of times.  Ref.~\onlinecite{GroverTurner} computed the topological entanglement entropy of the $d=3$ toric code using a different prescription, which produces the same result.

\section{Topological entanglement entropy of fracton models}
\label{sec:fracton}

In this section, we compute the topological entanglement entropy of fracton models.  We begin (Sec.~\ref{sec:decoupled_tc}) with a simple warm-up example, a stack of decoupled two-dimensional toric code layers, which is related to the X-cube model by a coupled-layer construction. \cite{Vijay2017, Hermele}  Then we proceed to consider the X-cube model (Sec.~\ref{sec:x-cube}) and Haah's code (Sec.~\ref{sec:haah}).
 
\subsection{Decoupled layers of $d=2$ toric codes \label{sec:decoupled_tc}}

We consider a stack of $d=2$ toric codes, with layers spaced uniformly and arranged normal to the $z$ axis. We consider the ``square torus'' PQWT prescription shown in Fig.~\ref{fig:ABCD_3d_2}, with topological entanglement entropy defined by Eq.~(\ref{eq:stopo-pqwt}).  The intersection of each layer with $P$ is the annulus-shaped region discussed in Sec.~\ref{sec:toric2d}, which has two non-local stabilizers.  The regions $Q$, $W$ and $T$ do not support non-local stabilizers, so the topological entanglement entropy is two bits per layer.
Therefore, choosing the unit of length to be the layer spacing, and choosing the regions $P$, $Q$, $W$, $T$ to contain precisely $R$ toric code layers, we obtain
\begin{equation}
S^{PQWT}_{\text{topo}} = - 2 R \text{.}
\end{equation}
In this example of decoupled layers, it is no surprise that we obtain a $R$-linear term in $S_{{\rm topo}}$.  This term is of interest because it also appears in fracton models.

We remark that slightly changing the detailed specificiation of the regions can alter the constant term in $S^{PQWT}_{{\rm topo}}$.  For instance, instead of stating the regions contain precisely $R$ layers, we could define them to extend a distance $R$ along the $z$-axis between toric code layers on the ``bottom'' and ``top'' surfaces, and to contain these surface layers.  Then each region intersects $R+1$ layers, and we obtain $S^{PQWT}_{{\rm topo}} = -2 R - 2$.  
Note that this changes the constant term by an even integer; it is possible that this term does have a robust meaning modulo two.  In a generic system in the same phase as the decoupled stack we are considering, we would not have precise control over the number of layers intersected by these regions, so that only the $R$-linear term is clearly meaningful.  This illustrates a general point that, in $d=3$, one should be cautious in ascribing any meaning to the constant term in $S_{{\rm topo}}$ when an $R$-linear term is present.

We also consider the topological entanglement entropy using the $d=3$ ABC prescription shown in Fig.~\ref{fig:ABCD_3d}, with the preferred axis along the $z$-axis.  Considering a single layer, this reduces to the $d=2$ ABC prescription of Fig.~\ref{fig:ABCD}, giving a topological entanglement entropy of one bit per layer.  Again choosing the regions to intersect precisely $R$ layers, we have
\begin{equation}
S^{ABC}_{{\rm topo}} = - R \text{.}
\end{equation}

The example of decoupled layers of $d=2$ toric codes is also instructive in that it illustrates the answer can depend on the {\it orientation} of the entanglement cut. Indeed, different results would be obtained for either the ABC or PQWT prescriptions, if we choose the preferred axis to lie in an arbitrary direction, since the number of intersecting layers would be different. Although this somewhat complicates the interpretation of $S_{{\rm topo}}$, it has the advantage of providing a means to identify the ``natural'' axes for entanglement in this system, by rotating the orientation of the preferred axis so as to obtain a maximal answer.  In fracton models, this could potentially help to discover new coupled-layer constructions along the lines of Refs.~\onlinecite{Vijay2017,Hermele}.

We also consider a stacking of decoupled toric code layers along the $x$, $y$ and $z$ directions simultaneously, as in the coupled-layer construction of the X-cube model.\cite{Vijay2017,Hermele}  In this case, we obtain the \emph{same} results for topological entropy using the two prescriptions employed above, because these prescriptions do not capture the topological entanglement of the layers normal to the $x$ and $y$ axes.  We are not aware of a single ABC or PQWT type prescription that captures all of the topological entanglement in this system in 
one shot.  Instead, it seems to be necessary to compute $S_{{\rm topo}}$ for different sets of regions to obtain a full picture of the non-local entanglement. A notion of `recoverable information' that can be defined for stabilizer codes offers a complementary perspective to the one in the present paper and may achieve this goal.

\subsection{X-cube model \label{sec:x-cube}}

We now apply the stabilizer formalism to compute our first new result: namely, the entanglement entropy of the X-cube model, an archetypal example of a `type I' fracton phase~\cite{Vijay2016}. The model is defined on a cubic lattice with a spin $1/2$ variable on each link, with Hamiltonian 
\begin{equation}
H_{\text{XC}} = - \sum_{v} \left(A_v^{(xy)} + A_v^{(yz)} + A_v^{(zx)}\right) - \sum_c B_c ,
\label{eq:HXC}
\end{equation}
where the $A$-type stabilizers involve a product of four $Z_i$ operators that surround a vertex in one of three orthogonal planes, and the $B$-type stabilizers involve a product of twelve $X_i$ operators around a elementary cube.

To compute the topological entanglement entropy, we employ the construction Eq. \eqref{eq:stopo-pqwt}, using the regions shown in Fig.~\ref{fig:ABCD_3d_2}, and taking the preferred axis to be the $z$-axis. As discussed in Sec.~\ref{sec:loc-nonloc}, $S^{PQWT}_{{\rm topo}}$ is given by counting non-local stabilizers via Eq.~(\ref{eqn:spqwt-nonlocal}).  It follows from Appendix~\ref{app:xcube-locgen} that regions $Q$, $W$ and $T$ have only local stabilizers (\emph{i.e.} their stabilizer groups are locally generated), so we have $S^{PQWT}_{{\rm topo}} = - \Omega_P$, with $\Omega_P$ the number of non-local stabilizers in $P$.

\begin{figure}[t]
\includegraphics[width=.3\textwidth]{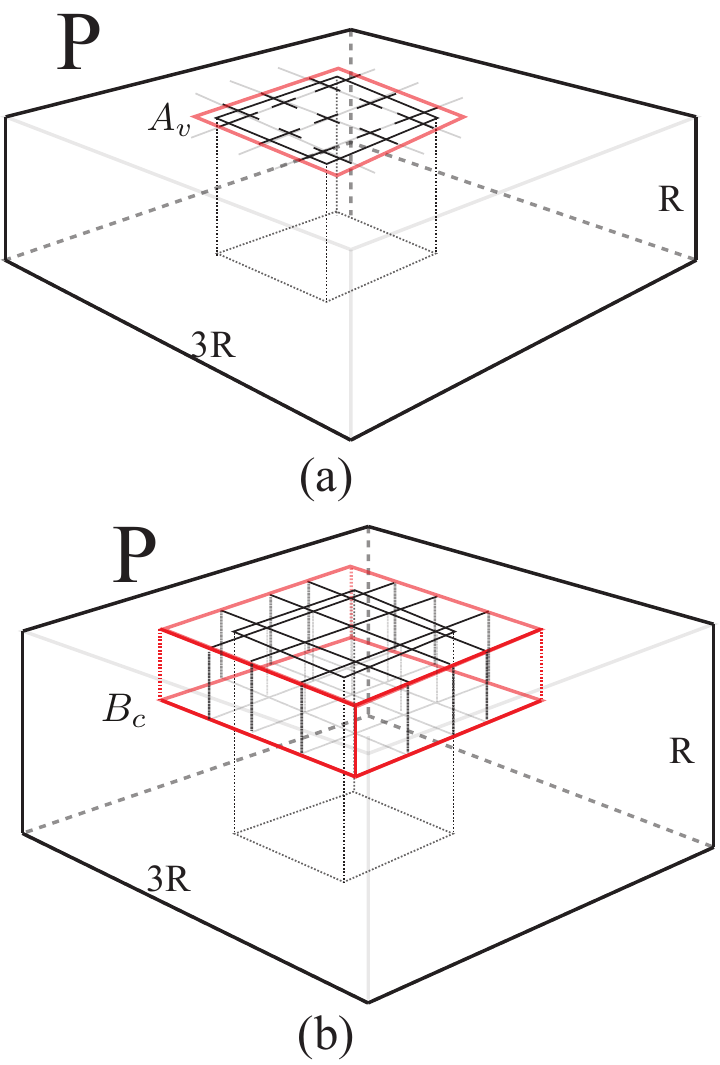}
\caption{(a)A non-local $Z$ stabilizer (red) of the X-cube model in region $P$ is constructed by taking a product of vertex terms over the ``hole'' in the center of $P$. (b) A non-local $X$ stabilizer (red) of the X-cube model in region $P$ is constructed by taking a product of cube terms over the ``hole'' in the center of $P$. \label{fig:operator_x_cube}}
\end{figure} 

To determine $\Omega_P = \Omega^X_P + \Omega^Z_P$, we first note that a product of cube terms over an $xy$ plane layer cutting through the ``hole'' in the center of $P$ produces a non-local $X$ stabilizer, as shown in Fig.~\ref{fig:operator_x_cube}.  There are $R$ different layers, and we expect that the resulting $R$ non-local stabilizers are independent in the sense that they cannot be deformed into one another by taking products with local stabilizers, so that $\Omega^X_P = R$.  The reason for this expectation is that each of these non-local stabilizers is a closed-loop string operator for a distinct non-trivial quasi-particle excitation confined to move in the corresponding $xy$ plane.\cite{Vijay2016,Hermele}  It should not be possible to change the particle type of a string operator by multiplying it with local operators.  We also verified  that  $\Omega^X_P = R$ using the numerical method of Sec.~\ref{sec:calculating}, for $R = 2,\dots,10$.   Similarly, taking a product of $xy$-plane ``vertex type'' stabilizers over the hole in $P$ also gives a non-local stabilizer.  There are $R+1$ such $xy$-plane layers, giving $\Omega^Z_P = R+1$, which we again verified numerically for $R = 2,\dots,8$.  Therefore we find $\Omega_P = 2R +1$ and
\begin{equation}
S^{PQWT}_{\text{topo}} = -2R-1 \text{.} \label{eq: xcstopo}
\end{equation}
We note that the $R$-linear term in this result is identical to that obtained in a stack of decoupled $d=2$ toric codes along $x$, $y$ and $z$ axes; as discussed in Sec.~\ref{sec:decoupled_tc}, only the layers normal to the $z$-axis contribute to the topological entanglement entropy for this choice of the regions $P$, $Q$, $W$, $T$.

As in the case of the toric code, the linear term of this result can be understood within a loop condensate picture.  In the $Z$ basis, configurations satisfying the vertex terms of the Hamiltonian can be viewed in terms of strings of flipped links $\ell$ with $Z_{\ell}=-1$, where in every $\{100\}$ plane the strings form closed loops.  Each $\{100\}$ plane thus gives a non-local contribution of $-c$ to $S_A$, where $c$ is the number of connected components in the intersection of the plane with the boundary of $A$.  Applying this simple rule to $S^{PQWT}_{{\rm topo}}$, we find that each $xy$ plane contributes $-2$, while $yz$ and $xz$ planes do not contribute; this reproduces the $-2 R$ term obtained above.  We note that this is the same loop condensate picture as for a stack of decoupled toric codes.  In the X-cube model, the layer-by-layer loop constraints in the ground state wave function are not truly independent; a more detailed analysis taking this into account would presumably also  reproduce the constant term in $S^{PQWT}_{{\rm topo}}$.

It is evident that, as for the case of decoupled layers of $d=2$ toric codes, the number of topological constraints (and hence the topological entanglement entropy) will depend on the {\it orientation} of the entanglement surface. Our discussion here is for an entanglement surface aligned with the symmetry axes of the problem. The entanglement entropy for arbitrary orientations could be evaluated using analogous methods, but we do not discuss it further here. 

We also compute the topological entanglement entropy using the ABC prescription, with regions shown in Fig.~\ref{fig:ABCD_3d}.  Because regions $A$, $B$, $C$, $AB$ and so on all have locally generated stabilizer groups generated by cube stabilizers and $xy$ and $xz$ plane vertex stabilizers (Appendix~\ref{app:xcube-locgen}), following Sec.~\ref{sec:loc-nonloc} and Appendix~\ref{app:nll}, $S^{ABC}_{{\rm topo}}$ is determined by counting local stabilizers whose support is split among all of $A$, $B$ and $C$.  The detailed geometry can be chosen so that this only occurs for cube stabilizers, and the number of these cube stabilizers is $R$.  Therefore we find
\begin{equation}
S^{ABC}_{{\rm topo}} = - R \text{,}
\end{equation}
where the vanishing of the constant term is presumably unimportant, because this prescription does not cancel all constant local contributions.  We observe that the coefficients of the topological entanglement entropy using both ABC and PQWT prescriptions in the X-cube model are exactly as in the corresponding system of decoupled toric codes.  Evidently, the linear term in the topological entanglement entropy is insensitive to the $m$-string condensation that occurs going from the decoupled toric codes to X-cube phase.\cite{Hermele}

\subsection{Haah's code \label{sec:haah}}
\begin{figure}[t]
\includegraphics[width=.45\textwidth]{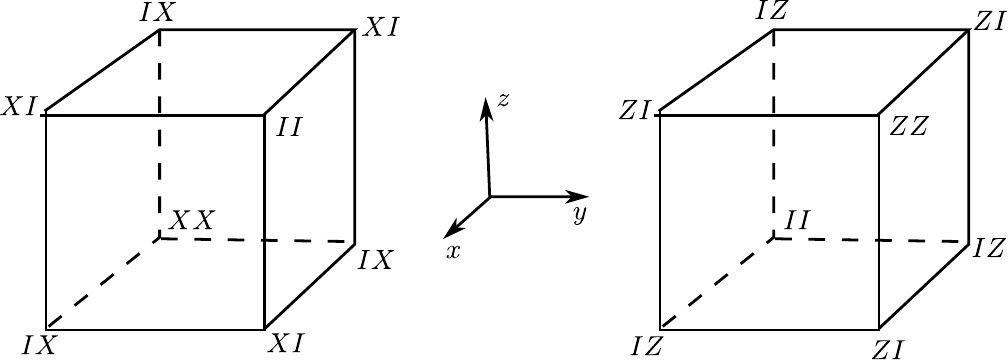}
\caption{$A_c$ (right) and $B_c$ (left) terms in Haah's code, showing our choice of coordinate axes. Each site has two spins. $X$ and $Z$ denote the corresponding Pauli operators. $I$ represents the  identity operator.\label{fig:Haah_code} }
\end{figure}

We now turn to Haah's code, the archetype of a `type-II' fracton model~\cite{Vijay2016}. This model is also defined on a cubic lattice, but now with two spin-$1/2$ variables on every {\it vertex}. The $Z$ and $X$ type stabilizer operators now consist of products of $Z$ and $X$  Pauli matrices around an elementary cube with a Hamiltonian of the form
\begin{equation}
H_{\textrm{Haah}} = - \sum_c( A_c + B_c),
\end{equation}
where $A_c$ and $B_c$ denote the products of $Z$ and $X$ Pauli operators around the vertices of a cube specified in Fig.~\ref{fig:Haah_code}.

To compute the topological entanglement entropy, we first employ the ABC prescription with regions as shown in Fig.~\ref{fig:ABCD_3d}.  It is shown in Appendix~\ref{app:haah-locgen} that the stabilizer groups for regions $A$, $B$, $C$, $AB$, and so on are all locally generated.  Therefore, following the discussion of Sec.~\ref{sec:loc-nonloc}, $- S^{ABC}_{{\rm topo}}$ is the number of local stabilizers contained in $ABC$ that have support split among all three regions $A$,$B$, $C$.  These stabilizers reside on cubes along the axis where the three regions meet, and there are $2(R-1)$ of them, where the factor of $2$ accounts for counting both $X$ and $Z$ stabilizers.  Therefore, 
\begin{equation}
S^{ABC}_{{\rm topo}} = - 2 R + 2 \text{,}
\end{equation}
where only the $R$-linear term is expected to have any universal meaning.

We also consider the topological entanglement entropy captured by two different PQWT prescriptions, determined using the numerical method of Sec.~\ref{sec:calculating}.  We first simplify the problem using the spatial inversion symmetry of Haah's code, which acts non-trivially on the spins, sending $X \to Z$ and $Z \to -X$, and also exchanging the two qubits on each site.  If a region $A$ is inversion-symmetric, then $\Omega^Z_A = \Omega^X_A$.  On the other hand, if two regions $A$ and $B$ are related to one another by inversion, then $\Omega^Z_A = \Omega^X_B$ and $\Omega^X_A = \Omega^Z_B$.  In the PQWT prescription of Fig.~\ref{fig:ABCD_3d_2}, regions $P$ and $T$ are inversion-symmetric, while inversion exchanges $Q$ and $W$.  This implies
\begin{equation}
S^{PQWT}_{{\rm topo}} = -2\Omega^Z_P  + 2\Omega^Z_Q + 2\Omega^Z_W - 2\Omega^Z_T \text{.} \label{eq:Stopoomega}
\end{equation}

Our numerical calculations lead to the conclusion that $S^{PQWT}_{{\rm topo}} = 0$ for the regions of Fig.~\ref{fig:ABCD_3d_2}. (In more detail, we show in Appendix~\ref{app:haah-locgen} that the stabilizer groups for regions $Q$, $W$ and $T$ are all locally generated, i.e. $ \Omega^Z_Q = \Omega^Z_W = \Omega^Z_T = 0$. Numerically, we find that $\Omega^Z_P=0$ for $R = 4, \dots, 11$.  While $\Omega^Z_P = 2$ for $R= 2$ and $\Omega^Z_P = 1$ for $R = 3$, this seems to be a finite-size effect.)  This result is strikingly different from $S^{ABC}_{{\rm topo}}$, while the linear term in these two entropies only differed by a factor of two for the X-cube model.  The contrast with the X-cube model suggests that Haah's code may not have a coupled-layer description where the layers lie in $\{100 \}$ planes.

\begin{figure}[t]
\includegraphics[width=0.98\columnwidth]{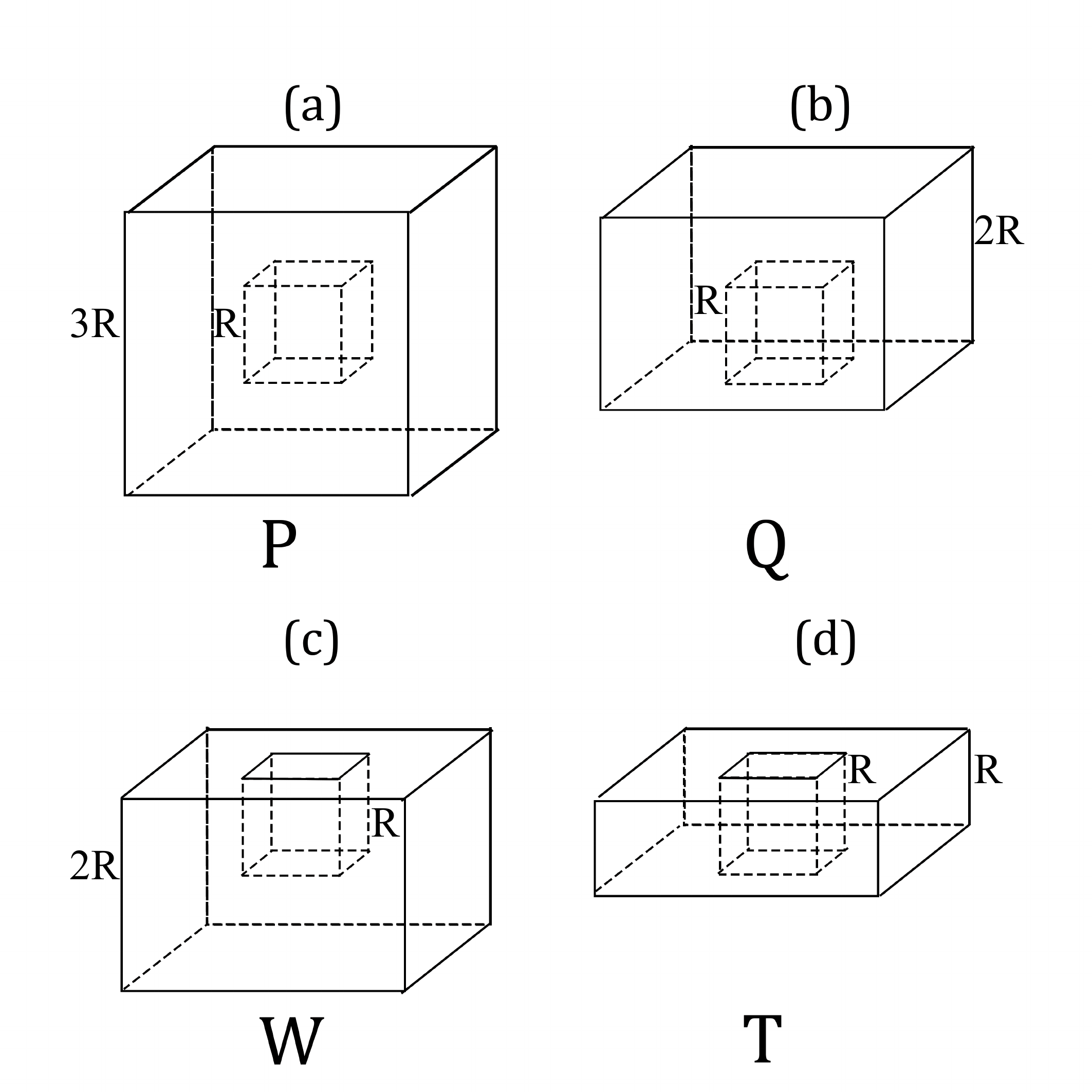}
\caption{Regions for an alternative PQWT prescription used for Haah's Code. Length of each edge is measured by the number of links. \label{fig:ABCDHaah}}
\end{figure}

To find a different PQWT prescription that does capture some of the non-local entanglement in Haah's code, we note that we should not expect the non-local stabilizers of Haah's code to be one-dimensional objects, as they are in the X-cube model. This expectation is based on the fact that none of the topologically charged excitations in Haah's code can be transported by string operators, so we should expect that any non-local stabilizers are higher-dimensional objects.  Moreover, this expectation is further substantiated by the fact that $\Omega^Z_P = 0$ for the solid torus region of Fig.~\ref{fig:ABCD_3d_2}. This motivates us to employ the PQWT prescription of Ref.~\onlinecite{CastelnovoChamon2008}, with regions shown in Fig.~\ref{fig:ABCDHaah}.  Here, the region $P$ is more isotropic, allowing for non-local stabilizers wrapping entirely around the interior cube.

The regions of  Fig.~\ref{fig:ABCDHaah} indeed give a non-zero result for $S^{PQWT}_{{\rm topo}}$.  Our numerical results are summarized in Table~\ref{tab:Haah_PQWT}, and we find
\begin{equation}
S^{PQWT}_{{\rm topo}} = -4 R + 12 \text{,}
\end{equation}
based on numerical calculations up through $R = 11$.  This topological entanglement entropy also has a $R$-linear term.  It is interesting to remark that, while Haah's code has a well-known intricate dependence of the ground state degeneracy on system size~\cite{haah2013commuting}, the behavior of the topological entanglement entropy is much simpler. 

\begin{table}[htbp!]
  \centering
\caption{The number of non-local stabilizers $\Omega^Z_{A}$ as a function of $R$ for regions $A = P,Q,W,T$ shown in Fig.~\ref{fig:ABCDHaah}.  The functional forms shown are exact from $R = 4$ up through $R = 11$ (the largest value of $R$ for which calculations were done).  These results determine $S^{PQWT}_{{\rm topo}}$ via Eq.~(\ref{eq:Stopoomega}).\label{tab:Haah_PQWT}}
\begin{tabular*}{\columnwidth}{@{\extracolsep{\fill}}|*5{>{\centering\arraybackslash}m{0.62in}|}  @{}m{0pt}@{}}
   \hline
$\Omega^Z_{P}$ & $ \Omega_{Q}^Z $ & $\Omega^Z_{W} $ &  $\Omega_{T}^Z$ & $S^{PQWT}_{\text{topo}}$  &\\ [2ex]
\hline
$6R - 7$  & $2R$ & $2R-1$ &$0$ & $-4 R + 12$  &\\ [2ex]
\hline
\end{tabular*}
\end{table}

At this point it is natural to ask if the topological entanglement entropy in Haah's code may also be given a geometrical interpretation in terms of constraints on the structure of the ground state wavefunction, like the loop condensate picture for toric code and X-cube models.  Because isolated fractons in Haah's code are created at corners of fractal operators, it seems likely that the ground state can be viewed as some kind of condensate of fractal objects. At present, to our knowledge there is no clearer meaning that can be given to this picture, or whether it can shed light on non-local entanglement.

\section{Localization-protected fracton order in excited states}

\label{sec:LPQO}
Thus far, we have focused exclusively on ground states. However, for {\it stabilizer} Hamiltonians, the entire spectrum shares the same entanglement entropy properties as the ground state (excited states differ only in their eigenvalues under stabilizer operators). The difference between ground states and excited states only manifests itself when the Hamiltonian is {\it perturbed} away from the stabilizer form. The entanglement structure of ground states is then `protected' from perturbations by the energy gap (as discussed above), whereas the excited states lack such protection. Upon perturbation, the excited states of translationally invariant fracton models are thus expected to {\it thermalize}~\cite{Prem2017} to {\it volume law} entanglement, in the process losing their topological order. Indeed, if one tries to construct the unitary $U$ of Eq.~(\ref{eq:localU}) using Schrieffer-Wolff perturbation theory outside the ground space, one finds that the perturbation theory diverges, suggesting that $U$ may not be a local unitary. Since our discussion thus far relied on the existence of a local unitary $U$, it has nothing further to say about the translationally invariant case.

However, as noted by Ref.~\onlinecite{LPQO}, this scenario could change dramatically once we break translational invariance by introducing quenched disorder: the topological order can be protected even in excited states by localization, in a manner that we now sketch. Consider  a {\it disordered} fracton Hamiltonian of the form
\begin{equation}
H = -\sum_i J^A_i A_i - \sum_j J^B_j B_j - \lambda H_{int},
\end{equation}
where the $J^A_i$ and $J^B_j$ are random numbers drawn from some distribution of width $W$. While the excitations are frozen (non-propagating) for {\it any} $W$ {\it at} the stabilizer point, once we add small perturbations they will be able to propagate for small $W$, but for sufficiently large $W$ the system can enter a {\it many body localized} phase~\cite{ARCMP}, where excitations cannot propagate freely. 
In this many-body localized phase, topological order (including fracton order) can persist at non-zero energy densities --- i.e., even in highly excited states. The challenge is how to detect this topological order. `Excited state degeneracy' cannot serve as a diagnostic, since the many body level spacing in the middle of the spectrum is exponentially small in the volume of the system and thus there is no longer a distinction between topological and other degeneracies in the thermodynamic limit. In Ref.~\onlinecite{LPQO}, non-local correlation functions --- related to the `Fredenhagen-Marcu' order parameters familiar to lattice gauge theorists --- were argued to be good diagnostics of topological order. These diagnostics were generalized to certain fractonic models in Ref.~\onlinecite{Devakul}, but nevertheless such non-local correlation functions can be challenging to compute. Here, we will demonstrate that topological entanglement can diagnose fracton topological order in excited states. 

The fact that excited states in the localized regime can support fracton topological order follows straightforwardly from our preceding discussions. First note that at $\lambda = 0$, excited states share the entanglement properties of the ground state (including topological entanglement), since the excited states are also eigenstates of the stabilizer operators. Now note that, in the localized phase, the unitary transformation $U$ is local, with at most exponential tails.
This follows because of the `mobility gap' in many body localized systems~\cite{NGH} i.e. the Schrieffer-Wolff perturbation theory has matrix elements in the numerator, and matrix elements vanish between near degenerate eigenstates.  Again, the topological entanglement entropy is non-local, and is expected to be unaffected by a local unitary transformation, subject to the same caveats discussed in the previous section.

We also note that the dressed stabilizers are simply the local integrals of motion or `lbits' of the localized system~\cite{Serbynlbits, lbits}, and these must be localized by postulate.  For a detailed construction of dressed integrals of motion via Schrieffer-Wolff perturbation theory, showing that these are local in the MBL regime, see Ref.~\onlinecite{Imbrie}. (For the cognoscenti, we note that our argument here parallels more closely the construction of l-bits via Wegner-Wilson flow in Ref.~\onlinecite{WegnerWilsonFlow}). 

We note that thus far we have assumed that a many body localized phase {\it can} exist in three dimensional lattice models. There is some debate about whether many body localization can arise in spatial dimensions $d > 1$ with random short range correlated disorder \cite{avalanches}, because of `thermalizing avalanches' triggered by rare regions. For truly random short range correlated disorder, our discussion applies to systems that are small enough to lack the relevant rare regions, and perhaps also in the thermodynamic limit, if the argument from Ref.~\onlinecite{avalanches, luitzavalanches, chandranavalanches} can be somehow circumvented. However, the problem may also be sidestepped by making the disorder long range correlated or {\it quasiperiodic}, such that the `rare region obstruction' identified in Ref.~\onlinecite{avalanches} does not apply. 

We therefore conclude that in disordered fracton models, fracton topological order can arise even in highly excited states, where it may be diagnosed through a `topological entanglement entropy' linear in the size of the subsystem. 

\section{Conclusions}
We have explicitly computed the entanglement entropy of two archetypal fracton models --- the X-cube model and Haah's code --- and have demonstrated the existence of a topological contribution to the entanglement entropy that is linear in the size of the subsystem. At a minimum, this provides a coarse characterization of fracton topological phases, in that for a given system, two states with distinct topological entanglement must be in distinct phases. There is also an obvious extension of this diagnostic to anisotropic models, wherein one separately considers the scaling of entanglement entropy with the size of subregion $A$ in the $x$, $y$ and $z$ directions respectively, thereby characterizing the topological entanglement with {\it three} indices. (More carefully, one would separately consider the topological entanglement entropy for a torus of thickness $R$ oriented in three orthogonal planes). In general the topological entanglement can depend not just on the size of the region, but also on its {\it orientation}, which may provide a useful means of diagnosing the symmetry axes of a phase by rotating the entanglement surface to obtain a maximal answer. 

What more information could be extracted from a study of entanglement? In two-dimensional topologically ordered phases, a careful analysis of the action of symmetries such as rotation and reflection within the manifold of degenerate ground states can provide insights into the fractionalized statistics of quasiparticle excitations in the phase; whether such manipulations can shed additional light on the properties of fracton excitations (that do not admit a quasiparticle description) remains an open question. It would also be interesting  to study  the {\it entanglement spectrum}~\cite{LiHaldane}, as this may contain more information than is encapsulated in entanglement entropy. Finally, the {\it dynamics} of entanglement has in other contexts (see, e.g. Refs.~\onlinecite{Prosen, BardarsonPollmanMoore}) provided much insight into the nature of thermalization and the approach to or avoidance of equilibrium. We leave investigation of these issues to future work.

\section*{Acknowledgements} M.H. and H.M. are supported by the U.S. Department of Energy, Office of Science, Basic Energy Sciences (BES) under Award number DE-SC0014415. SAP and RMN acknowledge support from the the Foundational Questions Institute (fqxi.org; grant no. FQXi-RFP-1617) through their fund at the Silicon Valley Community Foundation. This work is supported in part by the NSF under Grant No. DMR-1455366 (SAP).

{\bf Note added:} After this work was posted to the arXiv, we became aware of a parallel investigation \cite{he2017entanglement}
in which the entanglement entropy of ground states of fracton stabilizer codes was calculated, for certain bipartitions, using a completely different method. Where our results overlap, they agree. 

\begin{appendix}

\section{Regions with locally generated $G_A$ \label{app:enclosing}}
\label{app:locgen}

Here, we consider the X-cube model and Haah's code, and show that the group of stabilizers $G_A$ is locally generated for certain regions $A$ that we characterize.  Recall that in Sec.~\ref{sec:calculating}, we defined $G_A$ to be locally generated when $G_A = G_{A, loc}$, \emph{i.e.} when every stabilizer in $A$ is a product of local stabilizers.

\subsection{X-cube model}
\label{app:xcube-locgen}

\begin{figure}
\includegraphics[width=\columnwidth]{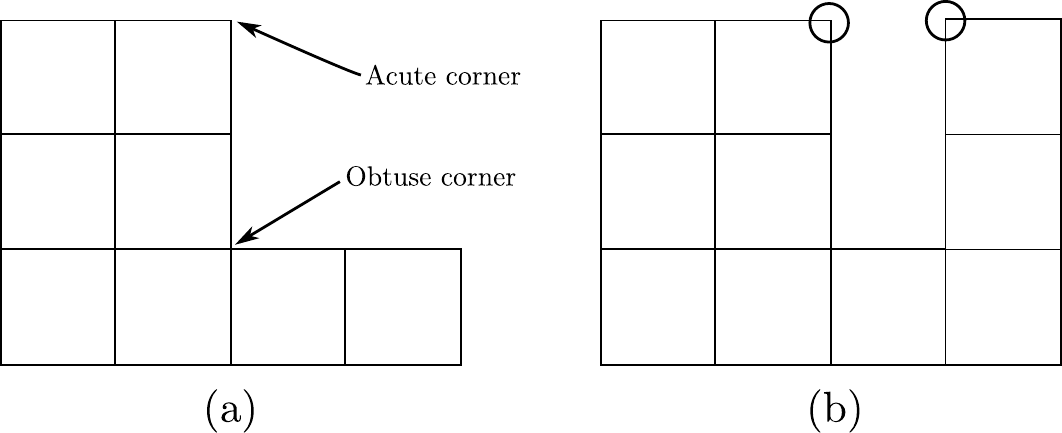}
\caption{Two square lattice slices that can be stacked to obtain a region $A$ of the cubic lattice.  Such slices have two kinds of corners, labeled acute and obsute in (a).  Our cleaning procedure goes through as long as acute corners are sufficiently isolated; region (a) satisfies this condition, while the circled corners in (b) cause the cleaning procedure for $Z$ stabilizers to fail for the corresponding region $A$.}
\label{fig:Xcubecleaning}
\end{figure}

In the X-cube model, we show that $G_A$ is locally generated for regions $A$ obtained by taking a simply connected region in the $xy$ plane, and stacking this region along the $z$-axis.  (The choice of plane and normal stacking direction is of course arbitrary, due to cubic symmetry.)  We assume the boundary of each $xy$ plane slice is a sequence of edges that are contained in $A$ and form a path in the cubic lattice.  Two examples of such regions for the $d=2$ square lattice are shown in Fig.~\ref{fig:Xcubecleaning}.  We further assume that acute corners of the slice (see Fig.~\ref{fig:Xcubecleaning}) are sufficiently isolated from other points on the boundary to carry out the cleaning procedure for $Z$ stabilizers discussed below.  A more precise statement of this assumption is given below; essentially, we are assuming the boundary of the slice is not too rough.  This condition allows the slice shown in Fig~\ref{fig:Xcubecleaning}a, but rules out that in Fig.~\ref{fig:Xcubecleaning}b. We refer to cubic lattice links oriented along the $x$, $y$ and $z$ axes as $x$, $y$ and $z$ links, respectively.  In addition to spins residing on $x$ and $y$ links contained in each slice, the region $A$ contains all $z$-links joining two adjacent slices.

In fact, under the assumptions given, we show an even stronger property: $G_A$  is locally generated using only $xy$-plane and $xz$-plane vertex stabilizers, and cube stabilizers.  (Alternatively, we can use $xy$-plane and $yz$-plane vertex stabilizers.)  Because there are no local constraints among $xy$ and $xz$ plane vertex stabilizers, using this as a generating set allows us to simply establish Eq.~(\ref{eq:stopo-abc}) in Appendix~\ref{app:nll}.

We first consider some operator ${\cal X}_0$ supported on $A$, which we take to be an $X$-stabilizer.  By definition, this means that ${\cal X}_0$ commutes with all local $Z$-stabilizers, including those not entirely supported on $A$.  We would like to show that ${\cal X}_0$ is a product of local $X$-stabilizers supported on $A$, which can be accomplished by a cleaning procedure, where we successively multiply ${\cal X}_0$ by such stabilizers until we obtain the identity operator.  We denote the $X$-stabilizer obtained from ${\cal X}_0$ at the current stage of the cleaning procedure by ${\cal X}$.

We begin with the bottom layer (\emph{i.e.} smallest $z$ coordinate) of ${\cal X}_0$.  In this slice, commutation between $xy$ plane vertex stabilizers is exactly as in the $d=2$ toric code.  Using the fact that the stacked region is simply connected, this implies that ${\cal X}$ restricted to this layer is a product of plaquette operators $\prod_{\ell \in p} X_{\ell}$, where $p$ is a square plaquette in the $xy$ plane.  We can thus clean this layer by multiplying ${\cal X}_0$ with a suitable product of local $X$ stabilizers whose cube centers lie just above the slice; this works because the restriction of these stabilizers to the slice are plaquette operators.

It appears this cleaning step may leave dangling $z$-links lying just above the bottom slice, where ${\cal X}|_{\ell} = X$ on such links.  The notation ${\cal X}|_\ell$ means the restriction of ${\cal X}$ to the link $\ell$, \emph{i.e.} ${\cal X}$ is a product over links of Pauli operators, and ${\cal X}|_\ell$ is the Pauli operator ($1$ or $X$) at the link $\ell$ appearing in the product.
By considering commutation of ${\cal X}$ with $xz$- and $yz$-plane vertex stabilizers whose centers lie in the original bottom layer, we see that ${\cal X}|_\ell = 1$ for all these dangling links.

These steps reduce the height of the stack by one unit cell, resulting in a new region of the with the same properties as the one we started with. Therefore, we can continue this procedure until ${\cal X}$ is only supported on a single slice.  We can show ${\cal X} = 1$ by considering commutation with $xz$ and $yz$ plane vertex stabilizers whose centers lie in the same slice.  Consider $x$-links in the slice of interest with some fixed $y$ coordinate.  Starting at large negative values of $x$, and moving in the positive $x$ direction, find the first link $\ell$ with ${\cal X}|_\ell = X$.  Then ${\cal X}$ anticommutes with the $xz$ vertex stabilizer whose center lies adjacent to this link in the negative $x$ direction.  This is a contradiction, and implies ${\cal X}|_\ell = 1$ for all $x$-links, and similarly for all $y$-links.  Therefore ${\cal X} = 1$ and we have reached the end of the cleaning process.  

Now we consider a $Z$ stabilizer ${\cal Z}_0$ supported on the same region $A$, and clean it via multiplication with vertex stabilizers.  Again we start with the bottom slice, where commutation between ${\cal Z}_0$ and cube stabilizers whose centers lie just below the slice is the same as in the $d=2$ toric code.  Therefore, ${\cal Z}_0$ restricted to this slice is a product of $xy$ plane vertex stabilizers, which we clean off.  This may leave dangling $z$-links just above the bottom slice, where ${\cal Z}|_{\ell} = Z$.  First consider a $z$-link $\ell$ just above an acute corner.  We assume there is a cube stabilizer that contains this link, but contains no other links of $A$; this is a more precise statement of our assumption that the acute corners are sufficiently isolated.  This cube stabilizer anticommutes with ${\cal Z}$ unless ${\cal Z}|_{\ell} = 1$.  All dangling $z$-links not above an acute corner can be cleaned by multiplying with a $xz$ plane vertex stabilizer, whose vertex lies just above the link to be cleaned.

This cleaning process can be continued until ${\cal Z}$ is supported on a bilayer consisting of the top slice, the slice just below it, and $z$-links joining these slices.  Both the top and bottom slice can be cleaned off as above, leaving only the $z$-links. Considering the set of $z$ links with ${\cal Z}|_\ell = Z$, we go to the leftmost column  of this layer (\emph{i.e.} smallest $x$), and find the $z$ link in this column with smallest $y$ coordinate.  We see that ${\cal Z}$ anticommutes with the cube stabilizer whose center lies in the same layer, and lies diagonally adjacent to the selected link in the negative $x$, negative $y$ direction.  This is a contradiction, so we must have ${\cal Z} = 1$, and the cleaning procedure is complete.

\subsection{Haah's code}
\label{app:haah-locgen}

\begin{figure}
\includegraphics[width=0.9\columnwidth]{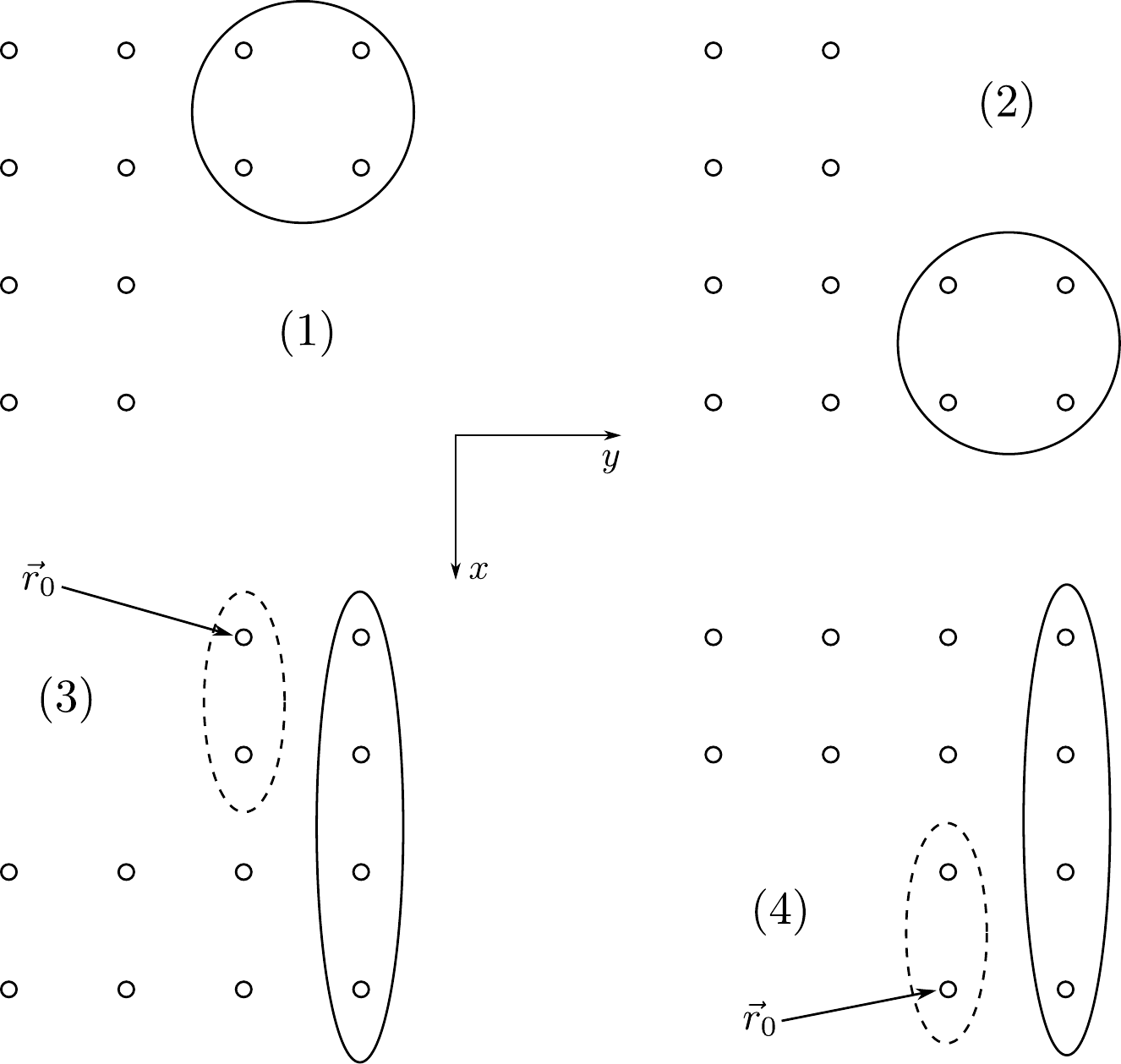}
\caption{Four types of L-shaped regions in Haah's code with axes shown.  Each region is a stack along the $z$-axis of the an $xy$ plane slice, with the top slice is shown for each region.  Regions of types $1, \dots, 4$ must be considered separately, because Haah's code lacks four-fold rotation symmetry about the $z$-axis.  The solid and dashed circled subregions, and the site label $\vr_0$ for the type 3 and 4 regions, are referred to in the text.}
\label{fig:Lshaped}
\end{figure}

In Haah's code, we first show that $G_A$ is locally generated for $A$ a rectangular prism, \emph{i.e.} a region containing all sites with $x$ coordinate satisfying $x_{{\rm min}} \leq x \leq x_{{\rm max}}$, and similarly for $y$ and $z$.  Axes are chosen as in Fig.~\ref{fig:Haah_code}. Next, we generalize this to show that $G_A$ is locally generated for certain ``L-shaped'' regions shown in Fig.~\ref{fig:Lshaped}.  In all cases, it is enough to concentrate on $Z$ stabilizers:  The rectangular prism regions are inversion-symmetric, so the corresponding result for $X$ stabilizers follow from inversion symmetry.  The L-shaped regions are not inversion-symmetric, but each region of one  type as shown in Fig.~\ref{fig:Lshaped} is related to a region of a different type under inversion, so that if all four types of regions have locally generated  $G^Z_A$, the corresponding statement for $X$ stabilizers follows.

Let $A$ be a rectangular prism region, and choose a $Z$ stabilizer ${\cal Z}$ supported in $A$.  We write
\begin{equation}
{\cal Z} = \prod_{\vec{r} \in A} \prod_{i = 0,1} (Z_{\vec{r},i})^{n(\vec{r},i)} \text{,}
\end{equation}
where $Z_{\vec{r}, i}$ is the $Z$ Pauli operator for the $i$th qubit at position $\vec{r} = (x,y,z)$, and where the choice of operator is specified by the binary numbers $n(\vec{r}, i) = 0,1$.  The restriction ${\cal Z}_{\vr}$  is given by
\begin{equation}
{\cal Z} |_{\vr} = (Z_{\vec{r},0})^{n(\vec{r},0)} (Z_{\vec{r},1})^{n(\vec{r},1)} \text{.}
\end{equation}
We often suppress some of the indices when using this notation, so for instance if $n(\vr,0) = n(\vr,1) = 1$, we write
\begin{equation}
{\cal Z} |_{\vr} = Z Z \text{.}
\end{equation}

In order for ${\cal Z}$ to be a stabilizer, it must commute with all the local $X$ stabilizers. We label local stabilizers $A_c$ and $B_c$ by the position of the corner with smallest values of the coordinates $x, y, z$, thus writing $A_c = A_{\vr}$ and $B_c = B_{\vr}$. Then the condition ${\cal Z} B_{\vec{r}} = B_{\vr} {\cal Z}$ can be written
\begin{eqnarray}
&& n(\vr, 0) + n(\vr, 1) + n(\vr + \hx,1) + n(\vr + \hy,1) \nonumber  \\ &+& n(\vr + \hz,1) + n(\vr + \hx + \hy,0)  + n(\vr + \hx + \hz,0) \nonumber \\
 &+& n(\vr + \hy + \hz,0)  = 0 \mod 2\text{.} \label{eqn:comm}
\end{eqnarray}
As discussed above for the X-cube model, we will carry out a cleaning process where we successively multiply ${\cal Z}$ by local $Z$ stabilizers contained in $A$ until ${\cal Z} = 1$.

\begin{figure}
\includegraphics[width=0.8\columnwidth]{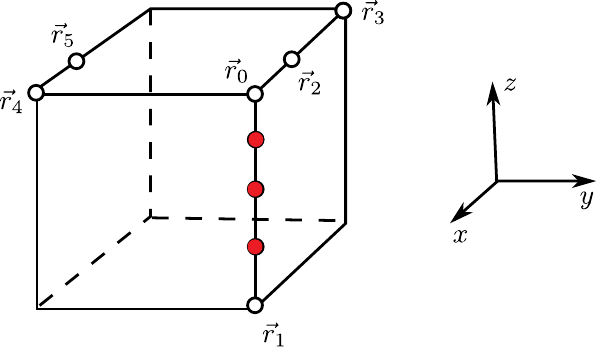}
\caption{Rectangular prism region $A$ in Haah's code, with some lattice sites (circles) labeled to facilitate the discussion of the cleaning procedure in the text.}
\label{fig:prism}
\end{figure}

Equation~(\ref{eqn:comm}) looks complicated, but it simplifies if we choose $\vr \to \vr_0$ to be the  corner of $A$ with largest $x$, $y$ and $z$ coordinates (see Fig.~\ref{fig:prism}), where it reduces to
\begin{equation}
n(\vr, 0) + n(\vr, 1) = 0 \mod 2 \text{.}
\end{equation}
This implies that either ${\cal Z}|_{\vr_0} = 1$ or ${\cal Z}|_{\vr_0} = ZZ$. Supposing the latter case, we  multiply ${\cal Z}$ by
$A_{\vr_0 - (\hx + \hy + \hz)}$, to obtain a new ${\cal Z}$ with ${\cal Z}|_{\vr_0} = 1$.

We  repeat this procedure, moving down in the $z$-direction, to clean operators in the column below $\vr_0$ (red dots in the figure) until $\vr_1$ is reached.  ${\cal Z}|_{\vr_1}$ cannot be cleaned in the same way, because the $Z$ stabilizer we would need to act with lies outside $A$.  However, again we have ${\cal Z}|_{\vr_1} = 1$ or ${\cal Z}|_{\vr_1} = ZZ$.  Considering  commutation of ${\cal Z}$ with $B_{\vr_1 - \hz}$, we find $n(\vr_1,1) = 0 \mod 2$.  Therefore we have ${\cal Z}|_{\vr_1} = 1$, and no cleaning is needed.

So far we have cleaned the vertical column below $\vr_0$.  To proceed, we move to $\vr_2$, and repeat the same procedure to clean the vertical column below $\vr_2$. Proceeding in this way we can clean until we reach the position $\vr_3$.  By the same reasoning as before, either ${\cal Z}|_{\vr_3} = 1$ or ${\cal Z}|_{\vr_3} = ZZ$.  We cannot apply the same cleaning procedure for $\vr_3$, because the local $Z$ stabilizer we would need to multiply lies outside of $A$.  However, commutation with $B_{\vr_3 - \hx}$ implies either ${\cal Z}|_{\vr_3} = 1$ or ${\cal Z}|_{\vr_3} = ZI$.  Therefore ${\cal Z}|_{\vr_3} = 1$ and no cleaning is needed.  The same argument applies to all the sites vertically below $\vr_3$, and therefore we have cleaned off the entire $+y$ face of ${\cal Z}$.

We can repeat the same steps to clean almost all of ${\cal Z}$ until position $\vr_4$ is reached.  Again we have either
${\cal Z}|_{\vr_4} = 1$ or ${\cal Z}|_{\vr_4} = ZZ$.  Commutation with $B_{\vr_4 - \hy}$ implies either
${\cal Z}|_{\vr_4} = 1$ or ${\cal Z}|_{\vr_4} = ZI$.  Therefore ${\cal Z}|_{\vr_4} = 1$ and no cleaning is needed.  We proceed vertically below $\vr_4$, and then on to position $\vr_5$, and so on, to see that ${\cal Z}|_{\vr} = 1$ everywhere on the remaining $-y$ face, and no further cleaning is needed.  This completes the argument, and we have shown $G_A$ is locally generated for $A$ a rectangular prism.

Now we consider L-shaped regions of four types, as shown in Fig.~\ref{fig:Lshaped}, and show that $G_A$ is locally generated for such regions. Regions of this geometry appear in the computation of $S_{{\rm topo}}$ via the ABC prescription (see Sec.~\ref{sec:haah}).  Each region is a stack along the $z$-axis for $z_{{\rm min}} \leq z \leq z_{{\rm max}}$ of an $xy$ plane slice.  The figure shows the top such slice ($z = z_{{\rm max}}$) for each region. Similar regions that are stacks along $x$ and $y$ axes can be obtained from these by three-fold rotational symmetry about the $[111]$ axis, and do not need to be considered separately.

For each type of region, we again consider a stabilizer ${\cal Z}$ supported in the region.  In each case, the entire subregion including and below the solid circled regions in the figure can be cleaned off by following the cleaning procedure described above for a rectangular prism region.  For type 1 and 2 regions, this results in a stabilizer ${\cal Z}$ supported within a rectangular prism, which we have already shown is a product of local $Z$ stabilizers.  For type 3 and 4 regions, it remains to consider the subregion including and below the dashed ovals.  Starting with the type 3 region, we consider the site $\vr_0$ in the top layer (see figure).  Considering commutation of ${\cal Z}$ with $B_{\vr_0 - \hx}$ and $B_{\vr_0 - \hx - \hy}$ implies ${\cal Z}|_{\vr_0} = 1$.  The same reasoning shows that ${\cal Z}|_{\vr} = 1$ for all sites in the column below $\vr_0$, and in the whole subregion containing and below the dashed oval.  This again reduces the problem to the already solved rectangular prism case.  The argument for type 4 regions proceeds essentially the same way, except that we consider commutation of ${\cal Z}$ with $B_{\vr_0 }$ and $B_{\vr_0 - \hy}$.

\begin{figure}
\includegraphics[width=0.9\columnwidth]{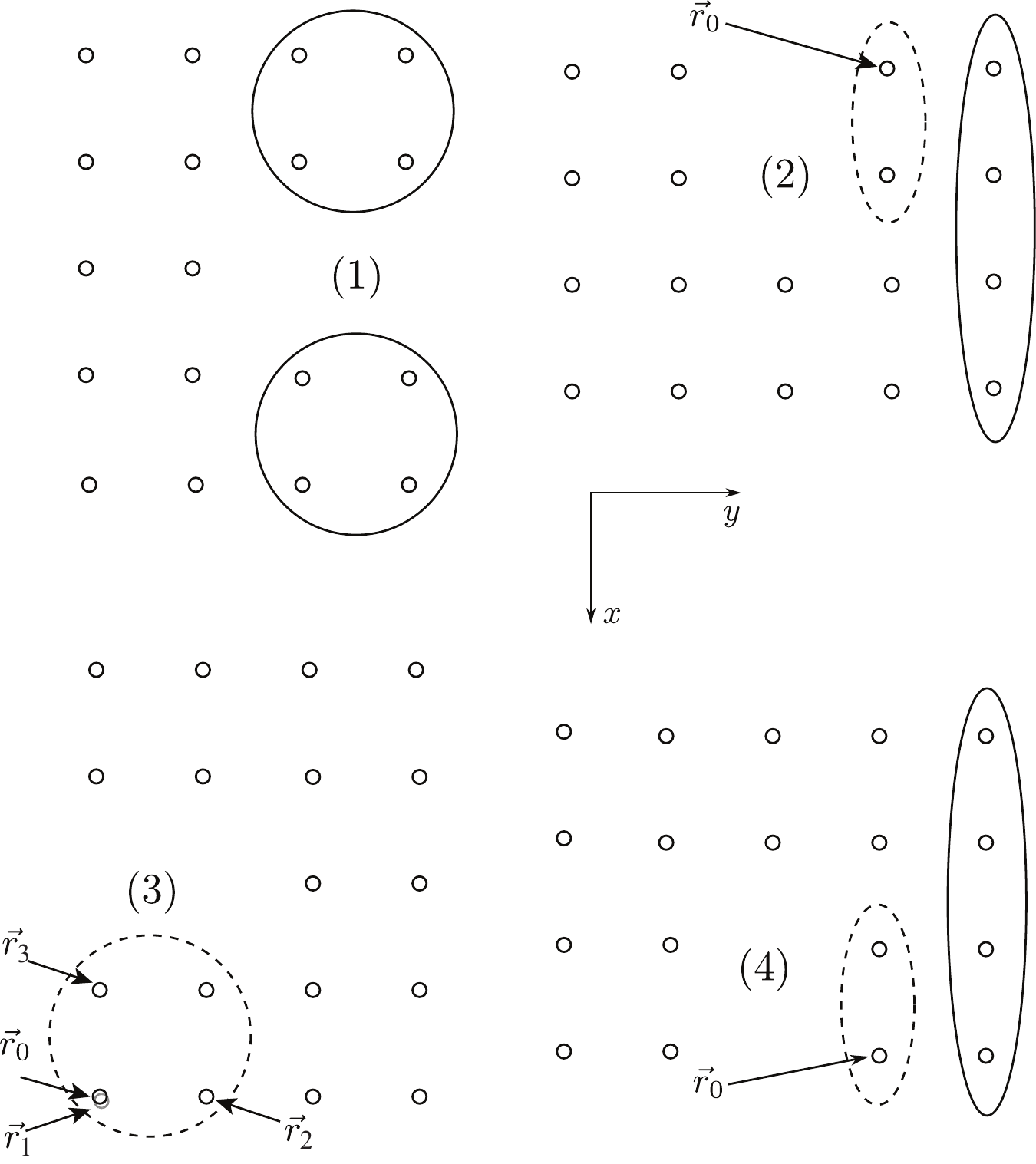}
\caption{Four types of C-shaped regions in Haah's code with axes shown.  Each region is a stack along the $z$-axis of the an $xy$ plane slice, with the top slice is shown for each region. }
\label{fig:Cshaped}
\end{figure}

Similarly, we can show that in the C-shaped regions involved in PQWT constructions like Fig. (\ref{fig:ABCD_3d_2}), $G_A$ is also locally generated. As shown in Fig.~(\ref{fig:Cshaped}), there are four possible C-shaped regions.  For each type of region, we still consider a stabilizer $\mathcal{Z}$ supported in that region. In each case, the entire subregion including and below the solid circled regions in the figure can be cleaned off by following the cleaning procedure described above for a rectangular prism region. For type 1 region, this results in a stabilizer ${\cal Z}$ supported within a rectangular prism, which we have already shown is a product of local $Z$ stabilizers.  For type 2 and 4 regions, the solid and dashed oval regions can be cleaned off in the same way as we clean the L-shaped regions. Then what is left are type 2 and 1 L-shaped regions respectively, shown in Fig.~\ref{fig:Lshaped}, which can be cleaned using the same procedure. For the type 3 region, it is more convenient to start from the lower left corner, i.e. $\vec{r}_0$. The commutation of $\mathcal{Z}$ with $B_{\vec{r}_0-\hy}$ leads to either $\mathcal{Z}_{\vr_0}=1$ or ${\cal Z}_{\vr_0} = Z I$. Suppose the latter one is true, then we can multiply by $A_{\vr_0-\hx-\hz}$ to clean this site. All sites below $\vr_0$ (with the same $x$ and $y$ coordinates) can be cleaned by a similar procedure, until one arrives at the bottom site $\vr_1$, which has $z=z_{min}$. The commutation of ${\cal Z}$ with both $B_{\vec{r}_1-\hy}$ and $B_{\vr_1-\hy-\hz}$ requires ${\cal Z}_{\vr_1}  =1$. The next column beginning with $\vr_2$ can be cleaned off in the same way. Next, we can move to $\vr_3$. Notice the commutation of ${\cal Z}$ with both $B_{\vec{r}_3-\hy}$ and $B_{\vr_3-\hx-\hy}$ implies that ${\cal Z}_{\vr_3}=1$. In a similar fashion we can proceed to clean the whole subregion in the dashed circle, resulting in a type 4 L-shaped region which we have already be able to clean  as discussed above.

\section{Simplified expressions for the topological entanglement entropies \label{app:nll}}

We derive the formula Eq.~(\ref{eqn:spqwt-nonlocal}) for $S^{PQWT}_{{\rm topo}}$ in terms of the number of non-local stabilizers in each region.  Treating $X$ and $Z$ stabilizers together, we pick a basis $B_{T, loc}$ for $G_{T, loc}$.  Because $T$ is a subset of all the other regions, we can extend this to bases $B_{Q,loc}, B_{W,loc}$ and $B_{P,loc}$ for $G_{Q,loc}, G_{W,loc}$ and $G_{P,loc}$.  Each of these bases can then be extended to a basis for the full stabilizer group in the corresponding region, including non-local stabilizers.  The topological entanglement entropy is
\begin{eqnarray}
S^{PQWT}_{{\rm topo}} &=& - \Omega_P + \Omega_Q + \Omega_W - \Omega_T  \\
&-& | B_{P,loc} | + | B_{Q,loc} | + | B_{W, loc} | - |B_{T,loc}| \text{.} \nonumber
\end{eqnarray}

To simplify this expression, we first note that
\begin{equation}
B_{T,loc} = B_{Q,loc} \cap B_{W, loc} \text{.}
\end{equation}
This holds because the local stabilizers added to extend $B_{T,loc}$ to $B_{Q,loc}$ are contained only in $Q$ and not in $W$, and vice versa.  Next we consider the union $B_{Q,loc} \cup B_{W, loc}$.  This set spans $G_{P,loc}$, because there is a basis for $G_{P,loc}$ where the basis elements are local operators that are thus fully contained either in $Q$ or $W$.  However it can happen that this set is not linearly independent.  Therefore we have
\begin{equation}
|B_{P,loc}| = |B_{Q,loc} \cup B_{W, loc}| - \Delta_{PQWT} \text{,}
\end{equation}
where $\Delta_{PQWT}$ can be thought of as the number of constraint equations obeyed by elements of $B_{Q,loc} \cup B_{W, loc}$.  These constraints must be non-local, involving stabilizers contained in $Q$ but not in $W$, and vice versa.  It follows that
 \begin{equation}
 S^{PQWT}_{{\rm topo}} = - \Omega_P + \Omega_Q + \Omega_W - \Omega_T + \Delta_{PQWT} \text{.}
 \end{equation}
 
We now discuss $\Delta_{PQWT}$ in more detail; among the models we consider, it can be non-zero only for the $d=3$ toric code.  Each constraint equation contributing to $\Delta_{PQWT}$ is a product of local stabilizers in $P$ that equals the identity operator.  Such a constraint can always be obtained as a product of \emph{local} constraints, involving stabilizers that may lie in a larger region containing $P$.  This can occur for the plaquette stabilizers  $d=3$ toric code. For example, using the regions $P,Q,W,T$ shown in Fig.~\ref{fig:ABCDHaah}, a product of plaquette stabilizers in $P$ over a surface $S$ enclosing the inner cube is the identity operator.  The plaquette stabilizers satisfy the local constraint that a product of stabilizers over the faces of an elementary cube is unity, and the non-local constraint in $P$ can be obtained by from these local constraints by taking a product over all cubes inside $S$.
 
However, there are no such non-local constraints in the other models we consider, where $\Delta_{PQWT} = 0$.  In the $d=2$ toric code and Haah's code, the stabilizers obey no local constraints.  This is also true for the cube stabilizers of the X-cube model.  The vertex stabilizers of the X-cube model do obey local constraints, but because each vertex stabilizer participates in exactly one local constraint, it is not possible to obtain non-local constraints contributing to $\Delta_{PQWT}$ by taking products of local ones.  

Now we give a similar discussion of $S^{ABC}_{{\rm topo}}$, to obtain the result that it is determined by stabilizers whose support is split among all of $A$, $B$ and $C$.  Here, we consider only models and regions where $A$, $B$ and $C$ are locally generated by a basis of local stabilizers that obey no local constraints.  This property is established for the X-cube model in Appendix~\ref{app:xcube-locgen} using a basis of cube stabilizers, and $xy$ and $xz$ plane vertex stabilizers.  For Haah's code and the $d=2$ toric code, the stabilizers obey no local constraints.  We choose bases $B_A$, $B_B$ and $B_C$ for the stabilizer groups $B_A$, $B_B$ and $B_C$, respectively.

To obtain a basis for the pairwise unions $AB$ and so on, we consider the set $b_{AB} = B_A \cup B_B$.  This set is clearly linearly independent, except possibly for the vertex stabilizers of the X-cube model.  Any linear relation would have to involve basis stabilizers in both $A$ and $B$.  It cannot be a non-local constraint as discussed above.  It also cannot be a local constraint near the boundary of $A$ and $B$, because the stabilizers in $B_A$ and $B_B$ are drawn from a subset of local stabilizers that do not obey any local constraints.  Therefore $b_{AB}$ is linearly independent.  In order to obtain a basis $B_{AB}$ for $G_{AB}$, we extend $b_{AB}$ by adding stabilizers $\delta_{AB}$ whose support is split between $A$ and $B$.

Finally we consider the union $ABC$.  To find a basis, we first consider the set $b_{ABC} = B_A \cup B_B \cup B_C \cup \delta_{AB} \cup \delta_{BC} \cup \delta_{AC}$.  Again, this set is linearly independent, and we extend it to a basis $B_{ABC}$ by adding a set of stabilizers $\delta_{ABC}$, whose support is split among the three regions.

We then obtain for the topological entanglement entropy
\begin{eqnarray}
S^{ABC}_{{\rm topo}} &=& - | B_A| - | B_B| - | B_C |  \nonumber \\
&+& | B_{AB} |  + | B_{BC} | + | B_{AC} | - | B_{ABC} | \nonumber \\
&=& - | \delta_{ABC} | \text{,}
\end{eqnarray}
the desired result.

\section{Schrieffer-Wolff perturbation theory \label{app:SW}}
In this section we discuss the construction of dressed stabilizers upon perturbation of the stabilizer Hamiltonian. The derivation is a variation on the standard method of Schrieffer-Wolff transformations, most closely related to the method of Wegner-Wilson flow discussed in Ref.~\onlinecite{WegnerWilsonFlow}. We outline it here mainly in the interests of completeness. 

Let the eigenvectors of the unperturbed Hamiltonian ($H_0$) be $\{\ket{n}\}$ with eigenvalues $\{E_n\}$. Let the new Hamiltonian be written as $H = H_0 + \lambda V$, where $V$ is some Hermitan operator and $\lambda$ is any real number. Also let the eigenvectors of this Hamiltonian be $\{\ket{n^\prime}\}$ with eigenvalues $\{E_n^\prime\}$. The idea is to find the unitary operator $U(\lambda)$ such that 
\begin{equation}
\ket{n^\prime} = U(\lambda) \ket{n}.
\end{equation}
If we assume that $U(\lambda)$ is an analytic function, then there exists a Hermitian operator-valued function $F(\lambda)$ such that 
\begin{equation}
U(\lambda) = \exp \left(i F(\lambda) \right).
\end{equation}
We may then expand $F(\lambda)$ as a power series in $\lambda$. To find a relation for $U_{mn} = \braket{m|n^\prime}$, note that 
\begin{equation}
\braket{m| H_0 + \lambda V|n^\prime} = E_m \braket{m|n^\prime} + \lambda \braket{m|V|n^\prime} =E_n^\prime \braket{m|n^\prime}.  
\end{equation}
We may  also expand the $V$ term in the original eigenbasis, to obtain the self-consistent equation
\begin{equation}
U_{mn}(\lambda) =\lambda \sum_k \frac{V_{mk}U_{kn} (\lambda)}{ E^\prime_n(\lambda)- E_m}. \label{eq:urecur}
\end{equation}
We now make the standard assumption that $V$ does not act within degenerate subspaces of $H_0$ (any portion of $V$ that does so act should be absorbed into our definition of $H_0$), so we only have to worry about `off diagonal' matrix elements of $V$. 
If we define
\begin{equation}
A(\lambda) = \sum_k \frac{V_{mk}U_{kn} (\lambda)}{ E^\prime_n(\lambda)- E_m} |m\rangle \langle n|,
\end{equation}
we can use \eqref{eq:urecur} to express the expansion coefficients as
\begin{equation}
\left( \frac{\partial^j U}{ \partial \lambda^j}\right)_{\lambda = 0} = j \left(\frac{\partial^{j-1} A}{ \partial \lambda^{j-1}} \right)_{\lambda=0}, 
\end{equation}
assuming the derivatives of $A$ are well behaved at $\lambda=0$. At first order, we have
\begin{equation}
\left( \frac{\partial U}{\partial \lambda}\right)_{\lambda=0}= i \left( \frac{\partial F}{\partial \lambda}\right)_{\lambda=0} = A_{\lambda = 0} = i L,
\end{equation}
where $L$ can be expressed in terms of its matrix elements in the old eigenbasis, 
\begin{equation}
L_{mn} = i \frac{V_{mn}}{E_m - E_n + 0},
\end{equation}
or $[H_0, L] = iV$. One can continue the expansion to any desired order. Note that if one is concerned only with the ground state, and the system is gapped within a topological sector, then the denominator has a non-zero lower bound.
 It may be shown using standard techniques that if the original Hamiltonian has a conserved quantity, ie. $[H_0, S] =0$, then one can find a conserved quantity for the perturbed Hamiltonian, $\tilde S$ by solving 
\begin{equation}
\frac{\partial \tilde{S}}{\partial \lambda} = i [L,\tilde{S}],
\end{equation}
The expert reader will recognize this as the equation of motion for Wegner-Wilson flow\cite{WegnerWilsonFlow}. It is well known that this sort of flow equation preserves locality of the integrals of motion both for gapped systems and for localized systems (see e.g. Refs.~\onlinecite{Imbrie, WegnerWilsonFlow} for recent discussions), where `local' means `local up to an exponentially decaying tail.'
\end{appendix}

 \bibliography{library}
 \end{document}